\documentclass[11pt, letterpaper]{article}

\usepackage{graphicx}
\usepackage[dvipsnames, svgnames, x11names]{xcolor}
\usepackage[margin=1in, top=1.2in, headheight=25pt]{geometry} 
\usepackage[hidelinks]{hyperref}
\usepackage{fancyhdr} 
\usepackage{tcolorbox} 
\usepackage[super,sort&compress]{natbib} 
\setcitestyle{citesep={,}}
\usepackage{float} 
\usepackage{ragged2e} 
\usepackage[version=4]{mhchem}
\usepackage{caption}   
\usepackage{booktabs} 
\usepackage{amssymb}   
\usepackage{xltabular}
\usepackage{siunitx}   
\usepackage{multirow} 
\usepackage{svg}

\usepackage[utf8]{inputenc}
\usepackage{textcomp}
\DeclareUnicodeCharacter{2212}{-} 
\DeclareUnicodeCharacter{2005}{\hspace{0.25em}} 
\DeclareUnicodeCharacter{2009}{\thinspace}      

\usepackage[sfdefault]{carlito} 

\definecolor{SBGold}{HTML}{EBB028}
\definecolor{SBBoxBackground}{HTML}{F8F8F8} 
\definecolor{SBDarkGrey}{HTML}{B3B3B3}

\newcommand{\CompanyLogoSmall}{\includegraphics[height=15pt]{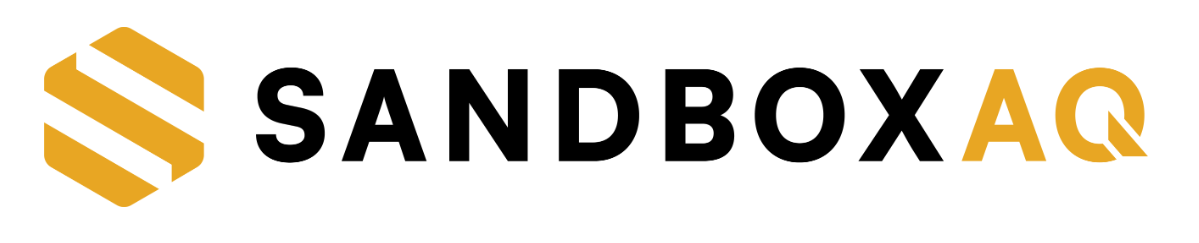}}

\newcommand{\PaperTitle}{AQVolt26: High-Temperature r$^2$SCAN Halide Dataset for Universal ML Potentials and Solid-State Batteries}
\newcommand{\sandboxaq}{SandboxAQ, Palo Alto, California, United States}
\newcommand{\nvidia}{Nvidia Corporation, Santa Clara, CA}
\newcommand{\correspondingsymbol}{*}        
\newcommand{\coauthorsymbol}{\textdagger}

\newcommand{\customauthors}{%
    \begin{flushleft}
        Jiyoon Kim\textsuperscript{1,\coauthorsymbol},
        Chuhong Wang\textsuperscript{1},
        Aayush R. Singh\textsuperscript{1},
        Tyler Sours\textsuperscript{1},
        Shivang Agarwal\textsuperscript{1},
        AJ Nish\textsuperscript{2},
        Paul Abruzzo\textsuperscript{2},
        Ang Xiao\textsuperscript{1},
        Omar Allam\textsuperscript{1,\coauthorsymbol,\correspondingsymbol}\\
        \vspace{1em}
        \textsuperscript{1}\sandboxaq \\
        \textsuperscript{2}\nvidia \\
        \vspace{0.5em}
        Correspondence: O.A. (omar.allam@sandboxaq.com)
    \end{flushleft}
 }

\newcommand{\AbstractText}{The demand for safe, high-energy-density batteries has spotlighted halide solid-state electrolytes, which offer the potential for enhanced ionic mobility, electrochemical stability, and interfacial deformability. Accelerating their discovery requires extensive molecular dynamics, which has been increasingly enabled by universal machine learning interatomic potentials trained on foundational datasets. However, the dynamic softness of halides poses a stringent test of whether general-purpose models can reliably replace first-principles calculations under the highly distorted, elevated-temperature regimes necessary to probe ion transport. Here, we present AQVolt26, a dataset of 322,656 r$^2$SCAN single-point calculations for lithium halides, generated via high-temperature configurational sampling across $\sim$5K structures. We demonstrate that foundational datasets provide a strong baseline for stable halide chemistries and transfer local forces well, however absolute energy predictions degrade in distorted higher-temperature regimes. Co-training with AQVolt26 resolves this blind spot. Furthermore, incorporating Materials Project relaxation data improves near-equilibrium performance but degrades extreme-strain robustness without enhancing high-temperature force accuracy. These results demonstrate that domain-specific configurational sampling is essential for the reliable dynamic screening of halide electrolytes. Furthermore, our findings suggest that while foundational models provide a robust base, they are most effective for dynamically soft solid-state chemistries when augmented with targeted, high-temperature data. Finally, we show that near-equilibrium relaxation data serves as a task-specific complement rather than a universally beneficial addition.}

\newcommand{\makeinfobox}{%
    \begin{tcolorbox}[
        colback=SBBoxBackground, 
        colframe=SBGold,          
        boxrule=0.0pt,
        arc=5mm,                  
        left=10mm, right=10mm, top=10mm, bottom=10mm, 
        boxsep=0pt, 
        width=\textwidth 
    ]
        {\raggedright\LARGE\bfseries\PaperTitle\par}
        \vspace{1em} 
        {\customauthors} 
        \vspace{0.5em} 
        \normalsize 
        \justify 
        \noindent\ignorespaces \AbstractText

        \vspace{1.5em}
        \noindent \raggedright \small
        \textbf{Models and Dataset:} \href{https://huggingface.co/collections/SandboxAQ/aqvolt26}{huggingface.co/collections/SandboxAQ/aqvolt26}
    \end{tcolorbox}
    \vspace{1em} 
}

\fancypagestyle{plain}{% 
    \fancyhf{} 
    \fancyhead[L]{\CompanyLogoSmall} 
    \fancyfoot[L]{\footnotesize\textsuperscript{\coauthorsymbol} Co-first authors; \textsuperscript{\correspondingsymbol}Corresponding author}
}

\begin{document}
\thispagestyle{plain} 
\makeinfobox

%%%%%%%%%%%%%%%%%%%%%%%%%%%%%%%%%%%%%%%%%%%%%%%%%%%%%%%%%%%%%%%%%%%%%%%%%%%%%%%%%%%%%%%%%%

%%%%%%%%%%%%%%%%%%%%%%%%%%%%%%%%%%%%%%%%%%%%%%%%%%%%%%%%%%%%%%%%%%%%%%%%%%%%%%%%%%%%%%%%%%

\section{Introduction}

\hspace{\parindent}The high demand of modern electronics has created an urgent need for advanced battery technologies that offer better capacity and longer lifespans \cite{Marom2011AMaterials, Xiao2021RecentBatteries, Nitta2015Li-ionFuture}. This is particularly critical for electric vehicles (EVs), which require batteries that combine high energy density, fast charging, long life cycles, and exceptional safety \cite{Lu2013AVehicles}. Lithium-ion batteries (LIBs) are currently the standard for EVs because lithium is lightweight and possesses a high electrochemical potential. These properties allow LIBs to achieve a superior energy density (265 Wh kg$^{-1}$) \cite{Placke2017LithiumDensity}, far outpacing older technologies like lead-acid (30-50 Wh kg$^{-1}$) or nickel-cadmium (45-80 Wh kg$^{-1}$) \cite{YekiniSuberu2014EnergyIntermittency}. However, LIBs rely on liquid electrolytes, which present significant risks regarding flammability, leakage, and thermal runaway caused by dendrites \cite{Lu2013AVehicles, Chen2021AdvancesViability, Cheng2017TowardReview, Yamada2019AdvancesElectrolytes}. To address these safety and energy density limitations, researchers are developing all-solid-state batteries (ASSBs). By replacing organic liquids with inorganic solid-state electrolytes (SSEs), ASSBs offer several key advantages: they eliminate flammability and leakage risks, in addition to acting as their own separator which frees up internal space and improves energy density relative to weight \cite{Banerjee2020InterfacesElectrolytes, Manthiram2017LithiumElectrolytes}. Despite their potential, ASSBs face hurdles, primarily because SSEs have lower ionic conductivity than liquids and struggle with mechanical and chemical stability at the electrode interfaces \cite{Takeda2016LithiumElectrolytes, Lim2020ABatteries}.

Halides represent one of the four primary categories of SSEs currently being researched alongside sulfides, oxides, and hydrides \cite{Asano2018SolidBatteries, Kanno1989IonicSystem, Lee2025Multi-Solid-ElectrolyteProspects}. Sulfides suffer from poor chemical and electrochemical stability \cite{Lau2018SulfideApplications}, oxides have relatively low ionic conductivity \cite{Subramanian2021ABatteries}, and hydrides are highly reducing electrolytes, which leaves halides to provide a third option with unique mechanical and chemical properties to address these disadvantages. Although halides are highly oxidative and may be challenging when paired with metallic electrodes \cite{Kwak2022EmergingApplications}, they can be valuable components in multi-solid-electrolyte systems, working in tandem with other electrolytes to prevent irreversible degradation with strategic placement at specific interfaces \cite{Wang2025CompositeOutlook, Hu2023Multi-layeredBatteries}. Research into SSEs relies heavily on electronic structure methods like Density Functional Theory (DFT) \cite{Kohn1965Self-ConsistentEffects}, because experimental methods often struggle to resolve atomic-scale mechanisms like ion hopping or interfacial reactions \cite{Leung2020DFTChallenges, Wu2025RecentReview, deKlerk2018AnalysisExample}.

DFT provides the gold standard for describing Potential Energy Surfaces (PES) but its computational costs scale with the cube of the electron count, making it often prohibitively expensive for simulating complex materials (e.g., amorphous solids, interfaces) or long-timescale phenomena like diffusion. Historically, researchers used classical force fields, which scale linearly with system size but sacrifice significant accuracy and are generally restricted to specific chemical bond types \cite{Lennard-Jones1931Cohesion, Daw1984Embedded-atomMetals, Senftle2016TheDirections}. In recent times, ML force fields have emerged to bridge this gap, using ML models to approximate the DFT PES with high efficiency \cite{Ko2023RecentPotentials}. Graph-based architectures are particularly effective, leveraging learned embedding vectors to handle diverse chemical compositions across the periodic table \cite{Chen2022ATable, Deng2023CHGNetModelling, Batatia2023MACE:Fields, Simeon2023TensorNet:Potentials}. Despite their promise, many foundation potentials often struggled with accuracy compared to custom-fitted models, primarily due to limitations in standard training datasets. For instance, a common source, the Materials Project relaxation dataset \cite{Jain2013Commentary:Innovation}, was designed for thermodynamic stability screening rather than learning force fields \cite{Riebesell2025APredictions}. For instance, it suffers from  discontinuities in the PES notably with forces and stresses through Perdew-Burke-Ernzerhof (PBE) and PBE+U functional mixing, and historical changes in computational settings that have introduced systematic noise into the dataset \cite{Shishkin2019DFT+UAdjustments, Qi2024RobustSampling, Deng2025SystematicPotentials}. 

Recently, foundational r$^2$SCAN-level datasets containing off-equilibrium states performed with consistent and higher fidelity electronic settings, such as MatPES \cite{Kaplan2025AMaterials} and MP-ALOE \cite{Kuner2025MP-ALOE:Potentials}, have been developed to address these issues and enable the development of highly capable universal ML force fields. However, it remains unclear whether models trained on these broad, general-purpose datasets can reliably substitute for ab initio methods during high-throughput screening of specific, complex material classes. For instance, soft crystalline lattices, such as halogenated SSEs, present a unique challenge for atomistic modeling. Their highly polarizable anions and frustrated structural arrangements create flattened, shallow potential energy basins \cite{Wood2021ParadigmsElectrolytes}. Consequently, ions undergo substantial excursions from their ideal crystallographic sites. Capturing this behavior is critical because the computational screening of solid electrolytes routinely employs elevated temperatures exceeding 1000 K to accelerate rare ion-hopping events to tractable timescales \cite{Chen2025High-throughputDescriptor}. To ensure models remain stable during these required dynamic simulations, the training dataset must comprehensively map the highly distorted, high-energy regions of the PES. Moreover, the choice of the r$^2$SCAN is particularly vital to capture mid-range dispersive interactions and the complex coordination environments inherent in soft halide lattices. 

\begin{figure}[ht!]
    \centering
    \includegraphics[scale=0.7, trim={3cm 3cm 4cm 3cm}, clip]{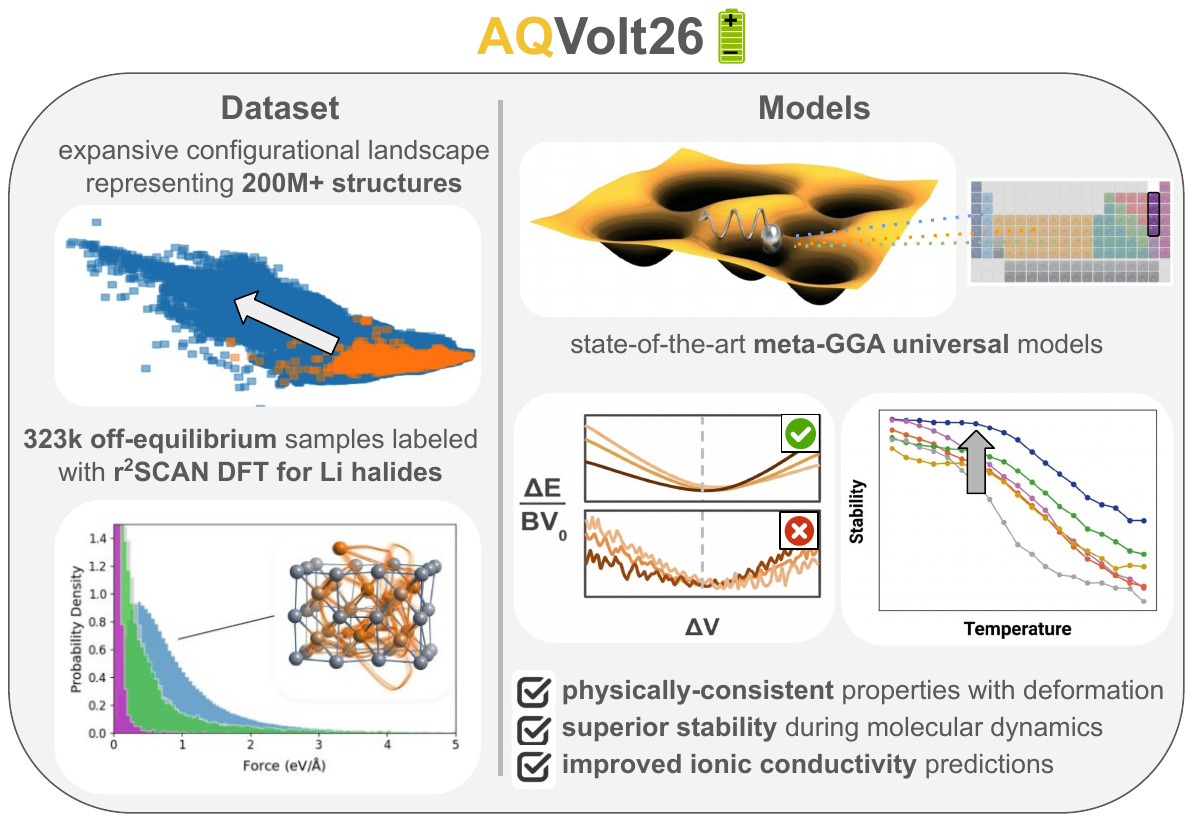}
    \caption{A summary of the AQVolt26 dataset and models. A configurational landscape of 200 million Li halide structures was selectively sampled and labeled with 322,656 r$^2$SCAN single-point calculations, generating the largest off-equilibrium dataset for solid-state electrolyte materials. Universal models were co-trained with foundational meta-GGA datasets using state-of-the-art architectures, demonstrating superior performance for materials with applied strain, stability during molecular dynamics simulations, and more accurate ionic conductivity values relative to experimentally-validated results.}
    \label{fig:overview}
\end{figure}

In this work, we present AQVolt26, a tailored dataset designed to address the dynamic modeling requirements of lithium halide SSEs (Figure \ref{fig:overview}). To efficiently capture this extreme configurational landscape, we generate millions of structures using surrogate-driven high-temperature phase space exploration, followed by strategic dimensionality reduction using the 2DIRECT sampling methodology developed by Qi et al. \cite{Qi2024RobustSampling}. We find that while models trained on foundational datasets perform adequately on familiar compositions at moderate temperatures, their energy predictions on novel halide compositions at elevated temperatures require substantial improvement, even when corresponding force predictions remain reasonable. Co-training with both domain-specific off-equilibrium data and near-equilibrium data provides this critical improvement and significantly extends the reliable operational domain of these potentials. Ultimately, these findings reaffirm that dataset composition and targeted configurational sampling are as consequential as model architecture in achieving the dynamic stability required to discover next-generation solid-state battery materials. 

%%%%%%%%%%%%%%%%%%%%%%%%%%%%%%%%%%%%%%%%%%%%%%%%%%%%%%%%%%%%%%%%%%%%%%%%%%%%%%%%%%%%%%%%%%

\section{Methods}

\subsection{Configuration Generation and Sampling}

\hspace{\parindent}A total of 1,549 lithium halides (Fluorine F, Chlorine Cl, Bromine Br, Iodine I) that do not contain oxygen, radioactive elements, and noble gases were sourced from the r$^2$SCAN MatPES dataset \cite{Kaplan2025AMaterials}. Lithium oxyhalides that exist in the Materials Project were deprioritized because this space has been explored more thoroughly \cite{Tanaka2023NewBatteries} and oxides are more susceptible to sluggish kinetics relative to sulfides and halides \cite{Lee2025Multi-Solid-ElectrolyteProspects,Liu2024TuningElectrolytes}. However, additional halide systems were generated among ``empty" compounds which do not previously contain a mobile species (e.g. Li), because it is hypothesized that these ions in oxides often exhibit stronger confinement to their equilibrium crystallographic sites and therefore suffer from reduced mobility \cite{Rong2015MaterialsStructures,Chen2023ZirconConduction,Kim2025EvaluatingCathodes}. From the 2025.06.09 version of the Materials Project, 3,783 non-lithiated, non-radioactive, and noble-gas-free systems with energy above hull values below 0.05 eV/atom, at least one halide anion, and a redox-active metal (Ti, V, Cr, Mn, Fe, Co, Ni, Cu, Nb, Mo, Ru, Ag, W, Re, Sb, Bi) that can accommodate Li atoms were selected. Systems that could not be adequately reduced based on oxidation state analysis methods outlined by Rutt et al. \cite{Rutt2022ExpandingCathodes} were discarded. Since additional Li sites were identified using an insertion algorithm that proposes sites in charge density minima \cite{Shen2020AMaterials}, candidates with open-sourced All-Electron Charge files (AECCAR) data \cite{Shen2022AMaterials} were evaluated with ML interatomic potentials to find the lowest energy lithiated configurations, producing 1,412 additional Li halides denoted as ``stuffed". Since a majority of these compounds contain fluorine and larger halogens enhance ionic conductivity due to increased anion screening \cite{Kwak2022EmergingApplications}, halide anion substitutions were performed for these compounds to better represent the less prevalent halogens. Structures with significant framework changes, as defined by the default parameters in the StructureMatcher algorithm from the Python Materials Genomics (pymatgen) library \cite{Ong2013PythonAnalysis}, and those with energy above hull values greater than 0.05 eV/atom were discarded based on phase diagram data available in the Materials Project database \cite{Gunter2012CommunityProject}. This augmented the material space by 1,950, resulting in a total of 4,911 Li halide candidates.

Next, over 200 million configurations were generated with a state-of-the-art r$^2$SCAN ML force field trained on the MatPES and MP-ALOE dataset \cite{Kuner2025MP-ALOE:Potentials}. NpT MD simulations were conducted at 1 atm for 50 ps per system with a time step of 1 fs at temperatures ranging from 300-1500K in increments of 300K, based on temperature ranges used in prior Li halide solid-state electrolyte screenings \cite{Chen2025High-throughputDescriptor}. Structure encoding was performed via a pre-trained M3GNet model \cite{Chen2022ATable}. In line with previous work, structural features were defined using the intermediate readout layer output, while atomic environments were represented by the node features post-graph convolution. A two-stage principal component analysis and clustering process (2-stage DImensionality-Reduced Encoded Clusters with sTratified, 2DIRECT) that operates first on the structural space and subsequently on the atomic space was applied \cite{Qi2024RobustSampling}, maximizing the diversity of sampled structures while significantly reducing the computational load for first-principles calculations. From the 200 million, 8.5 million structures were effectively encoded, resulting in 1.1 billion structural features and 9.1 billion atomic features. BIRCH clustering was performed with a threshold of 0.1 in the structural and atomic feature space, effectively sampling 2\% of the configurational landscape with 186,627 structures.

\subsection{Density Functional Theory Calculations}

\hspace{\parindent}Single-point calculations were carried out using version 6.4.3 of the Vienna ab initio Simulation Package (VASP). The VASP input parameters were consistent with ``MatPESStaticSet" documented in pymatgen \cite{Ong2013PythonAnalysis}, to conduct regularized-restored Strongly Constrained and Appropriately Normed (r$^2$SCAN) meta-Generalized Gradient Approximation (GGA) simulations \cite{Furness2020AccurateApproximation}. However, due to a lack of pseudopotentials specific to r$^2$SCAN in VASP, the Perdew-Burke-Ernzerhof \cite{Perdew1996GeneralizedSimple} ``PBE64" library was used, as is common practice \cite{Horton2025AcceleratedProject}. Roughly 79\% of the single-point calculations converged within the provisioned compute resources, which is consistent with typical success rates with r$^2$SCAN \cite{Kaplan2025AMaterials}. NpT Spin-polarized AIMD simulations at the r$^2$SCAN level of theory were performed with the same convergence criteria as the single-point calculations at 1 atm, with a time step of 2 fs with the exception of 0.5 fs for hydrogen-containing materials. These AIMD calculations were performed for less than a picosecond per system from a subset of Li halide systems devoid of oxygen in the MatPES dataset, including halide anion substitutions following the protocol described above, for benchmarking and model training purposes. This approach supplemented the AQVolt26 dataset with 175,065 DFT labels. First-principles calculations were conducted on n1-standard-32 virtual machines on the Google Cloud Platform to eliminate numerical discrepancies with differing floating-point units and hardware architectures.

\subsection{Machine Learning Model Training}

\hspace{\parindent}The models developed and released in this work utilize the equivariant Smooth Energy Network (eSEN) architecture \cite{Fu2025LearningPrediction}. We adopt the multi-stage training protocol established by the eSEN developers and standardized in recent benchmarks \cite{Fu2025LearningPrediction, Wood2026UMA:Atoms, Sahoo2025TheInterfaces, Levine2026TheModels} (Table \ref{tab:hyperparameters}).

In the first stage, the model was pre-trained to directly predict atomic forces using an explicit regression head. This initial phase utilized bfloat16 precision and a restricted graph connectivity capped at 30 nearest neighbors. The initial learning rate was set to $8 \times 10^{-4}$ with a cosine annealing schedule. In the second stage, the direct-force regression head was discarded, and the model was fine-tuned to predict conservative forces obtained via the analytical gradient of the predicted energy with respect to the atomic coordinates. The training precision was increased to 32-bit floating-point (FP32). The learning rate during this conservative fine-tuning phase was reduced to $4 \times 10^{-4}$. Throughout both stages, we used the AdamW optimizer with a weight decay of $10^{-3}$ and applied gradient clipping. The network architecture consisted of 4 interaction layers. This model is designated as eSEN-L4. Further details can be found in Table \ref{tab:hyperparameters}. 

To reduce data redundancy, the MatPES dataset was combined with MP-ALOE structures that were not sourced from the Materials Project. Consistent with the MatPES dataset threshold for acceptable forces, DFT labels with maximum force magnitudes greater than 800 eV/\AA{} were excluded from training. Configurations with positive cohesive energies were omitted as well because they were broadly deemed to be physically unrealistic. However, we posit that their inclusion could be beneficial for explicitly mapping the highly repulsive walls of the PES. Provided these extreme states possess sound energy and force labels that do not destabilize the model fit, they could serve as valuable physical guardrails during molecular dynamics simulations. We leave the systematic evaluation and integration of these high-energy configurations for future investigations. A 90-5-5 train-validation-test split was applied. Model training was conducted on H100 GPUs on the NVIDIA DGX compute cluster.

\subsection{Model Benchmarking}

\hspace{\parindent}Three open-source datasets with r$^2$SCAN data were used to perform benchmarking across publicly-available models and the AQVolt26 dataset: MatPES \cite{Kaplan2025AMaterials}, MP-ALOE \cite{Kuner2025MP-ALOE:Potentials}, Materials Project \cite{Huang2025Cross-functionalPotentials}. The VASP settings used to produce these datasets are largely consistent. MatPES and AQVolt26 are comprised of r$^2$SCAN single-point calculations, in contrast to the Materials Project and MP-ALOE datasets which performed relaxations. A distinction between the latter two is that the MP-ALOE relaxation trajectories consist of only three ionic steps to quickly equilibrate structures with unknown lattice parameters \cite{Kuner2025MP-ALOE:Potentials}. Since the orbitals were initialized with PBE to accelerate convergence for MatPES and MP-ALOE, the exchange-correlation energy was approximated with both GGA and meta-GGA for these datasets. Although AQVolt26 data was produced without PBE GGA orbital initialization, the calculated energy, forces, and stress differences were negligible in the final r$^2$SCAN result (mean absolute errors on the order of 10$^{-7}$ eV/atom, 10$^{-3}$ eV/Å, 10$^{-2}$ kbar). Pretrained foundational MLIPs were used to benchmark on each dataset with the test split generated during AQVolt26 model training, because the exact training, validation, and test seed splits from publicly-available sources were unknown.

A ``near-equilibrium" benchmark task was carried out on a random subset of 1,000 structures from the Materials Project GNoME r$^2$SCAN recompute dataset \cite{Merchant2023ScalingDiscovery}. These systems were randomly perturbed with atomic displacements of 0.1 Å then relaxed with each model using a force convergence threshold of 0.05 eV/Å as well as the FIRE optimizer \cite{Bitzek2006StructuralSimple}. Structural fidelity was quantified by comparing the local environments of the optimized ML force field structures against their DFT-relaxed references. This was calculated as the Euclidean distance between their CrystalNN fingerprint vectors \cite{Zimmermann2020LocalSimilarity}, with shorter distances suggesting a result closer to the ab initio truth. 

MD stability benchmarks were run on a Li halides as well as a halide-free set, curated from the GNoME structure pool \cite{Merchant2023ScalingDiscovery} using the 2DIRECT method \cite{Qi2024RobustSampling} to evaluate how each model performs on unfamiliar chemistries and configurations. Each material was subjected to a heating ramp from 300K to 2,100K at 1 bar in the NpT ensemble for 50 ps with a time step of 1 fs. Three independent MD runs were recorded per structure to capture structural failure, which was defined as either the loss of atoms or a cell volume change exceeding 50\% of the initial volume. Furthermore, a ``holdout" set was created from all 703 Li halides in GNoME \cite{Merchant2023ScalingDiscovery} by running MLMD at 1300K for 25 ps to generate approximately 200k frames, selecting samples with the same 2DIRECT \cite{Qi2024RobustSampling} procedure used previously (Figure \ref{fig:gnome_holdout_sampling}), and performing r$^2$SCAN single-point calculations on the most diverse configurations, which resulted in 3,841 converged calculations (Figure \ref{fig:distribution_gnome}).

% MD stability benchmarks were run on experimental Li solid electrolyte structures sourced from the OBELiX dataset \cite{Therrien2025OBELiX:Electrolytes}. The 98 systems from OBELiX that could transform into ordered crystalline structures using pymatgen \cite{Ong2013PythonAnalysis} were prioritized. 

NpT MD simulations were also conducted on 9 ordered Li halide and 63 non-halide structures from OBELiX \cite{Therrien2025OBELiX:Electrolytes} to validate ionic conductivity predictions from ML interatomic potentials relative to experimentally-measured values. Supercell structures with a minimum length of 8 Å for the shortest lattice vector were generated and simulated at temperatures between 300K and 700K in increments of 100K for at least 110 ps, ignoring the first 10 ps to compute the mean square displacement of Li. The ionic conductivity was calculated according to the Nernst-Einstein relation, linearly fitting the mean square displacement as a function of time \cite{He2018StatisticalSimulations} as implemented in pymatgen \cite{Ong2013PythonAnalysis, Ong2013PhaseConductors, Mo2012FirstMaterial}.

%%%%%%%%%%%%%%%%%%%%%%%%%%%%%%%%%%%%%%%%%%%%%%%%%%%%%%%%%%%%%%%%%%%%%%%%%%%%%%%%%%%%%%%%%%

\section{Results}

\subsection{Dataset}

\begin{figure}[ht!]
    \centering
    \includegraphics[scale=0.35]{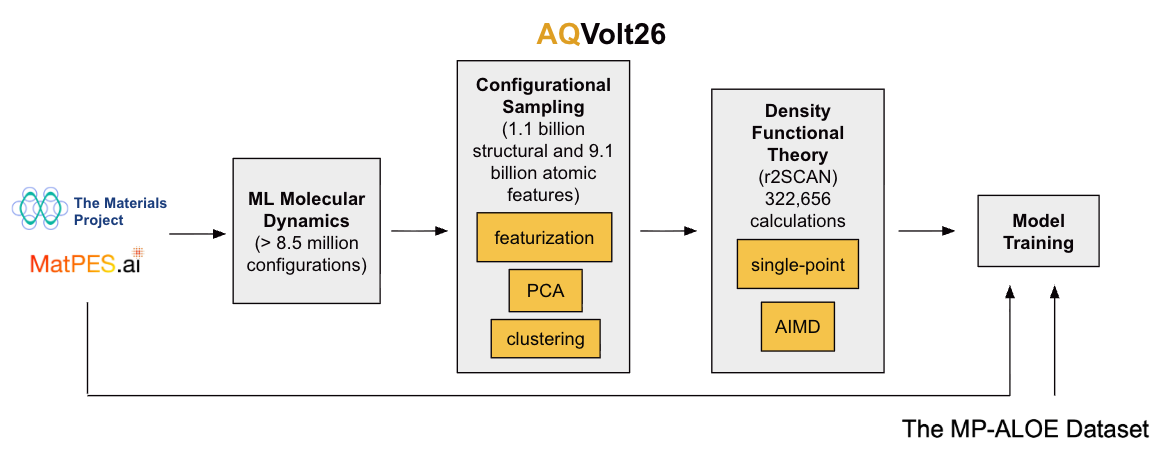}
    \includegraphics[scale=0.152]{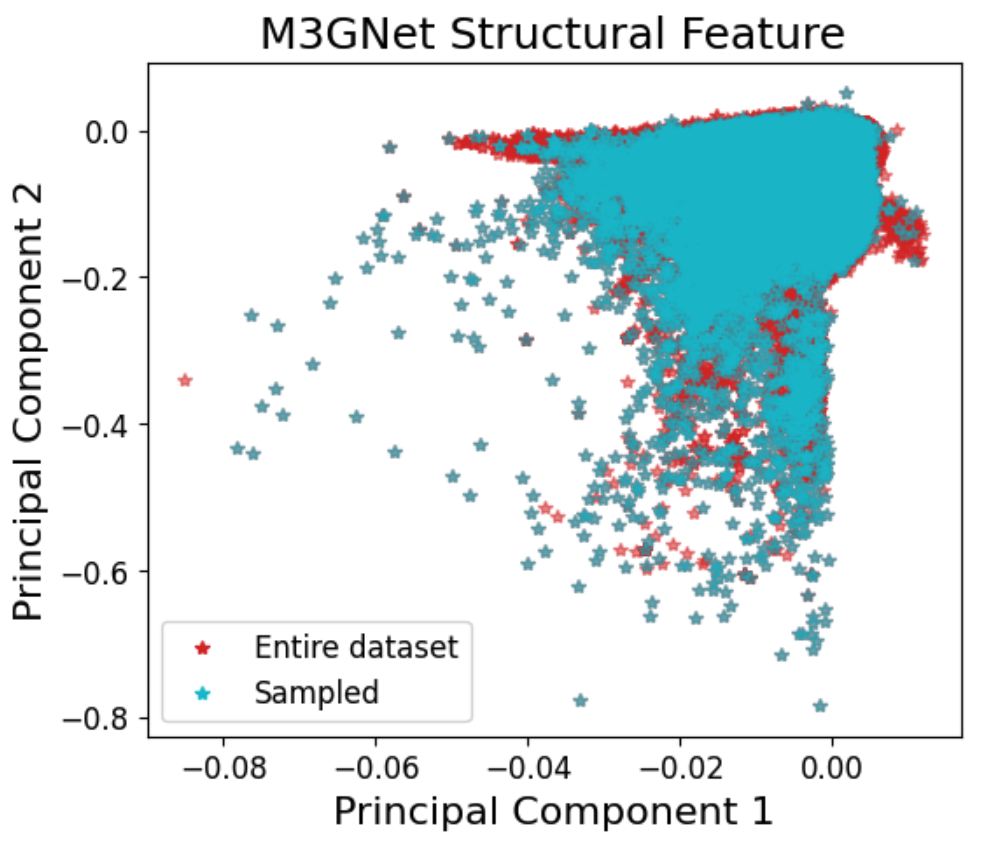}
    \includegraphics[scale=0.152]{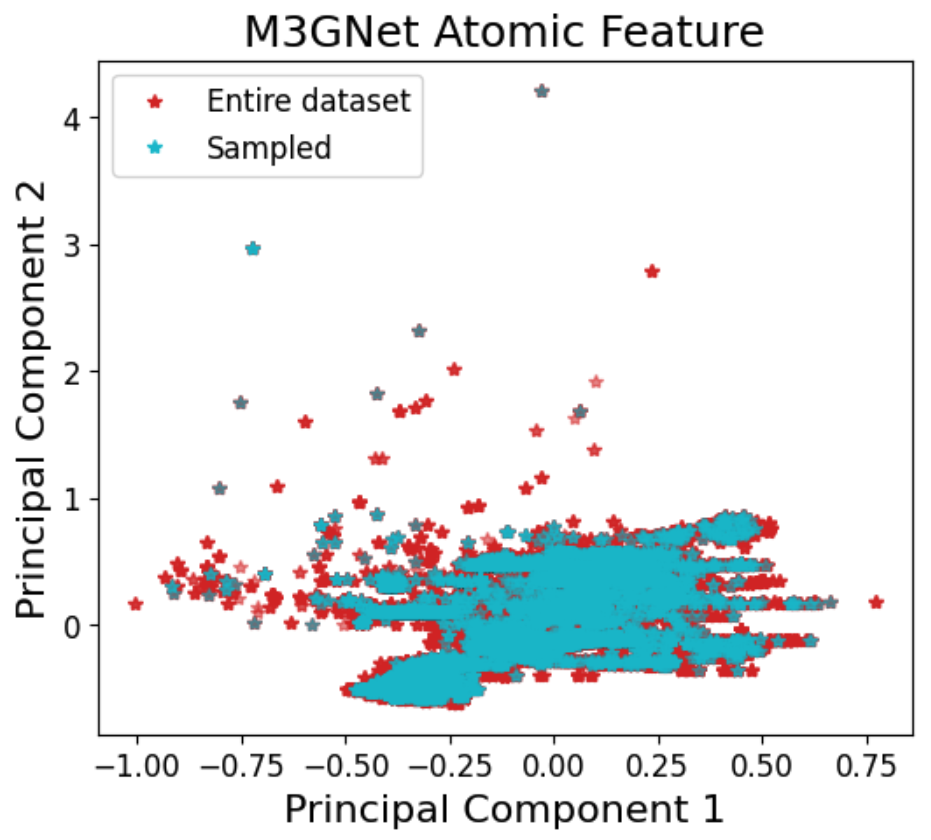}
    \includegraphics[scale=0.18]{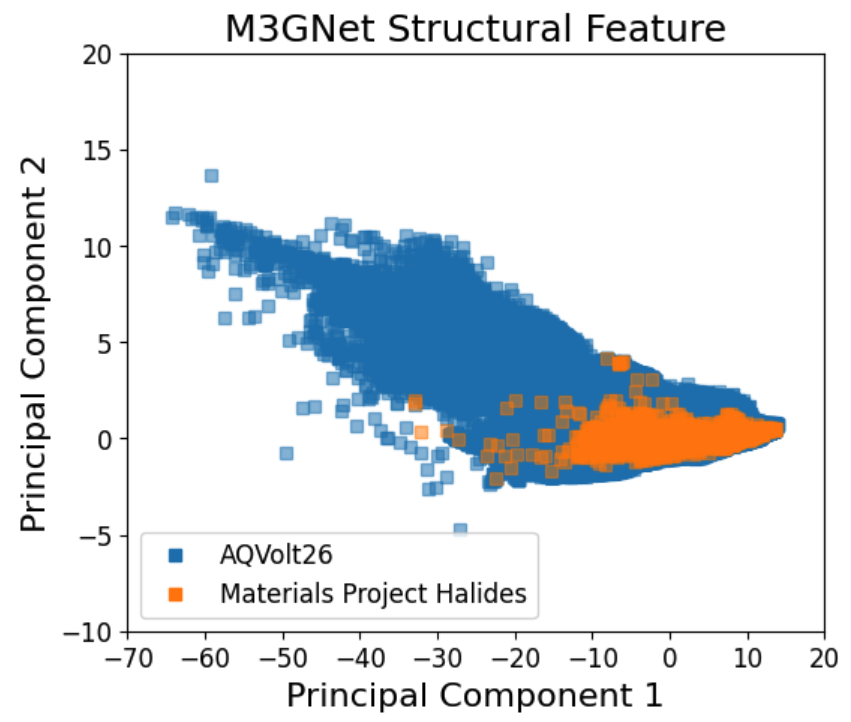}
    \caption{Overview of the AQVolt26 data generation and training approach (top). A dataset of 322,656 r$^2$SCAN single-point calculations was created through configurational generation with surrogate-driven phase space exploration (bottom left, in red) and dimensionality reduction with the 2DIRECT method \cite{Qi2024RobustSampling} (bottom left, in teal), covering a larger feature space compared to halides in the Materials Project (bottom right).}
    \label{fig:workflow}
\end{figure}

\hspace{\parindent}The AQVolt26 dataset is derived from a selection of 4,911 Li halide structures sourced from the Materials Project \cite{Huang2025Cross-functionalPotentials} and MatPES \cite{Kaplan2025AMaterials}, in addition to several novel structures generated as detailed in the Methods section (Figure \ref{fig:workflow}). To focus on systems highly relevant for SSE applications, compounds containing radioactive elements or inert noble gases were excluded (Figure \ref{fig:periodic_table}). In currently available datasets computed at the r$^2$SCAN level of theory, there are 5,640, 2,581, and 2,166 Li halide calculations in MP-ALOE \cite{Kuner2025MP-ALOE:Potentials}, MatPES \cite{Kaplan2025AMaterials}, and the Materials Project (MP) \cite{Huang2025Cross-functionalPotentials}, respectively. AQVolt26 exhibits significantly greater structural diversity, containing 4,336 unique halide structures, compared to 2,067 in MatPES, 1,560 in MP-ALOE, and 248 in the Materials Project. This disparity is expected, especially given that the Materials Project data primarily consists of near-equilibrium relaxations, and the expansive MP-ALOE dataset is derived from a narrower pool of structure optimizations. Consequently, AQVolt26 provides an unprecedented collection of 322,656 r$^2$SCAN DFT calculations for Li halide systems with comprehensive coverage of the periodic table. 

While existing datasets rarely sample lithium-containing halide materials, AQVolt26 places a heavy emphasis on these systems. Oxides are also relatively well-represented with 43,200 configurations, notably derived from novel lithium oxyhalide materials. The dominant transition metals in AQVolt26 are Manganese (Mn), Vanadium (V), and Iron (Fe).
%, reflecting a strategic exploration of earth-abundant, cost-effective metals for next-generation electrolytes, in contrast to the standard Cobalt and Nickel chemistries prevalent in LIBs. 
Furthermore, AQVolt26 features a balanced population of chlorides, bromides, and iodides, avoiding the over-representation of fluorides common in other datasets (Figure \ref{fig:periodic_table}). Strategic sampling using the 2DIRECT method \cite{Qi2024RobustSampling} enabled this vast structural and atomic feature space coverage using only 2\% of the originally generated configurations. Given that roughly 5M CPU-hours were used for compute for AQVolt26, this sampling approach effectively saved millions in operational costs.

\begin{figure}[ht!]
    \centering
    \includegraphics[scale=0.175]{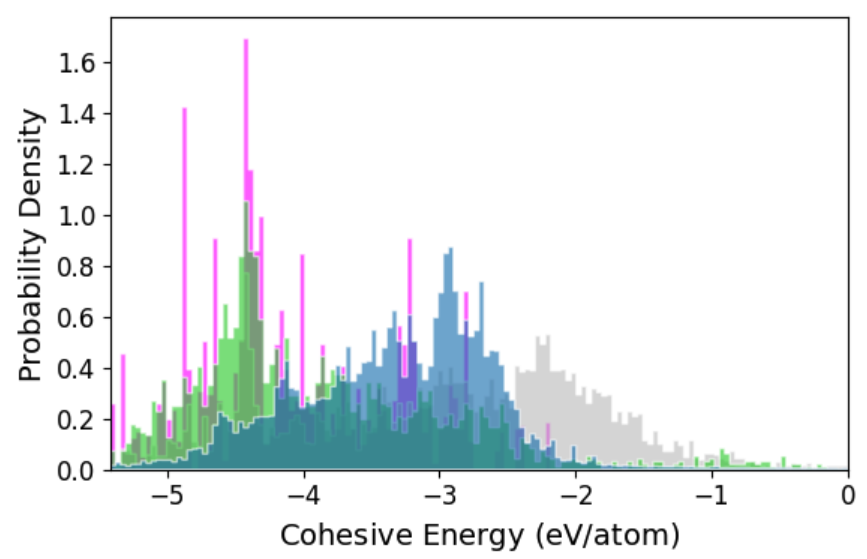}
    \includegraphics[scale=0.175]{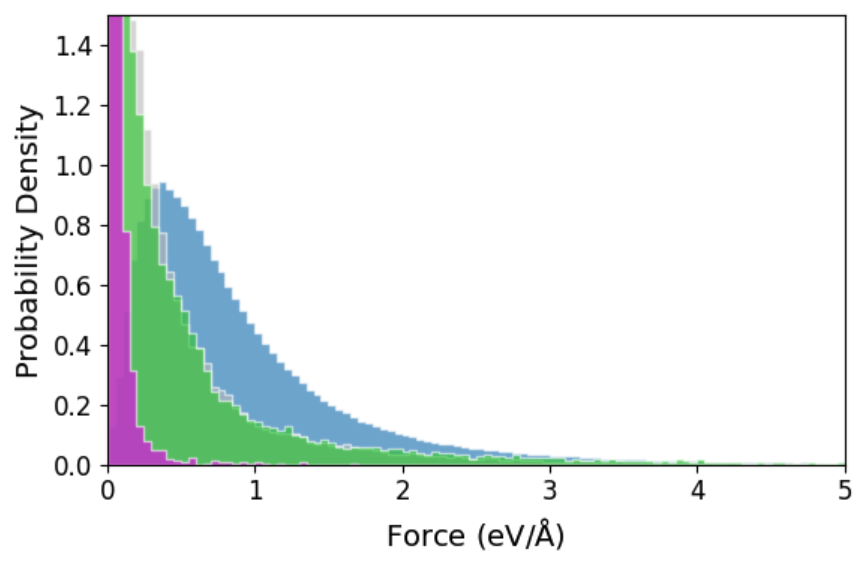}
    \includegraphics[scale=0.175]{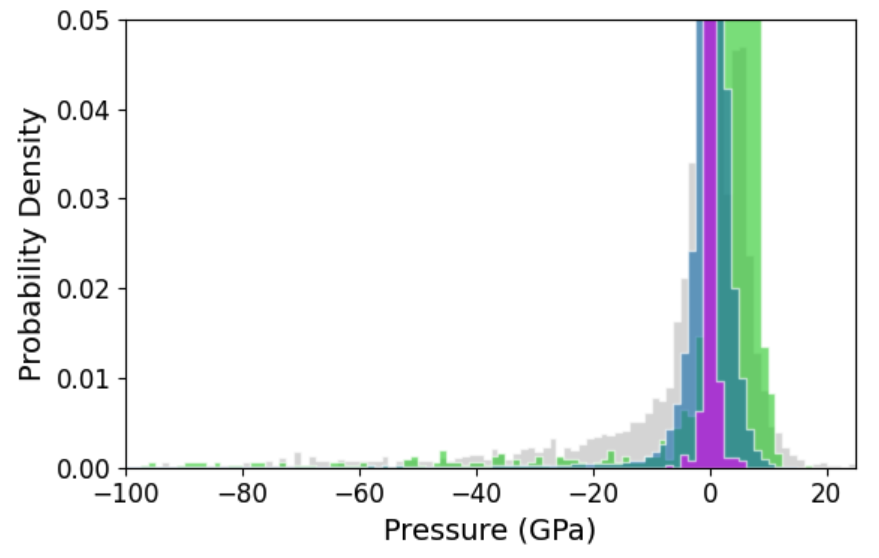}
    \includegraphics[scale=0.2]{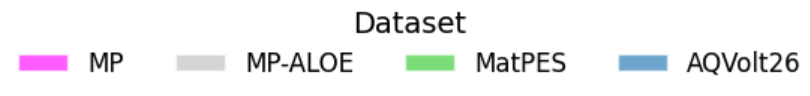}
    \caption{Comparison of the distributions of cohesive energies, interatomic force magnitudes, and pressures across four r$^2$SCAN datasets: AQVolt26 (blue), MatPES (green), MP-ALOE (grey), and Materials Project (pink) with 322,656, 2,581, 5,640, and 2,166 Li halide structures, respectively. AQVolt26 and MatPES configurations are primarily derived from single-point calculations at temperatures $\geq$ 300 K, while MP-ALOE and the Materials Project consist of structure optimizations at 0 K.}
    \label{fig:distribution}
\end{figure}

Of the four r$^2$SCAN datasets, AQVolt26 and MP-ALOE sample significantly more positive cohesive energies compared to MP and MatPES (Table \ref{tab:dataset_metrics}). The cohesive energy serves as a useful metric for identifying unphysical bonding or extreme states. While MP samples near-equilibrium states and contains zero flagged calculations, the high-temperature sampling in AQVolt26 frequently accesses configurations past the melting temperatures of halide systems, resulting in a broader tail in the interatomic force magnitude distribution (Figure \ref{fig:distribution}). Additionally, despite the focus on halogenated systems, AQVolt26 contains more structures with three or more elements and configurations exceeding ten atomic sites compared to MatPES, MP-ALOE, and MPr$^2$SCAN (Figure \ref{fig:element_site_distribution}). This aggressive sampling of the high-energy configurational landscape effectively captures the stochastic diversity of atomic environments necessary for modeling complex, multi-component SSEs.

\subsection{Model Performance}

\subsubsection{Fidelity Across Force Regimes}

\begin{figure}[ht!]
    \centering
    \includegraphics[width=0.95\linewidth]
    {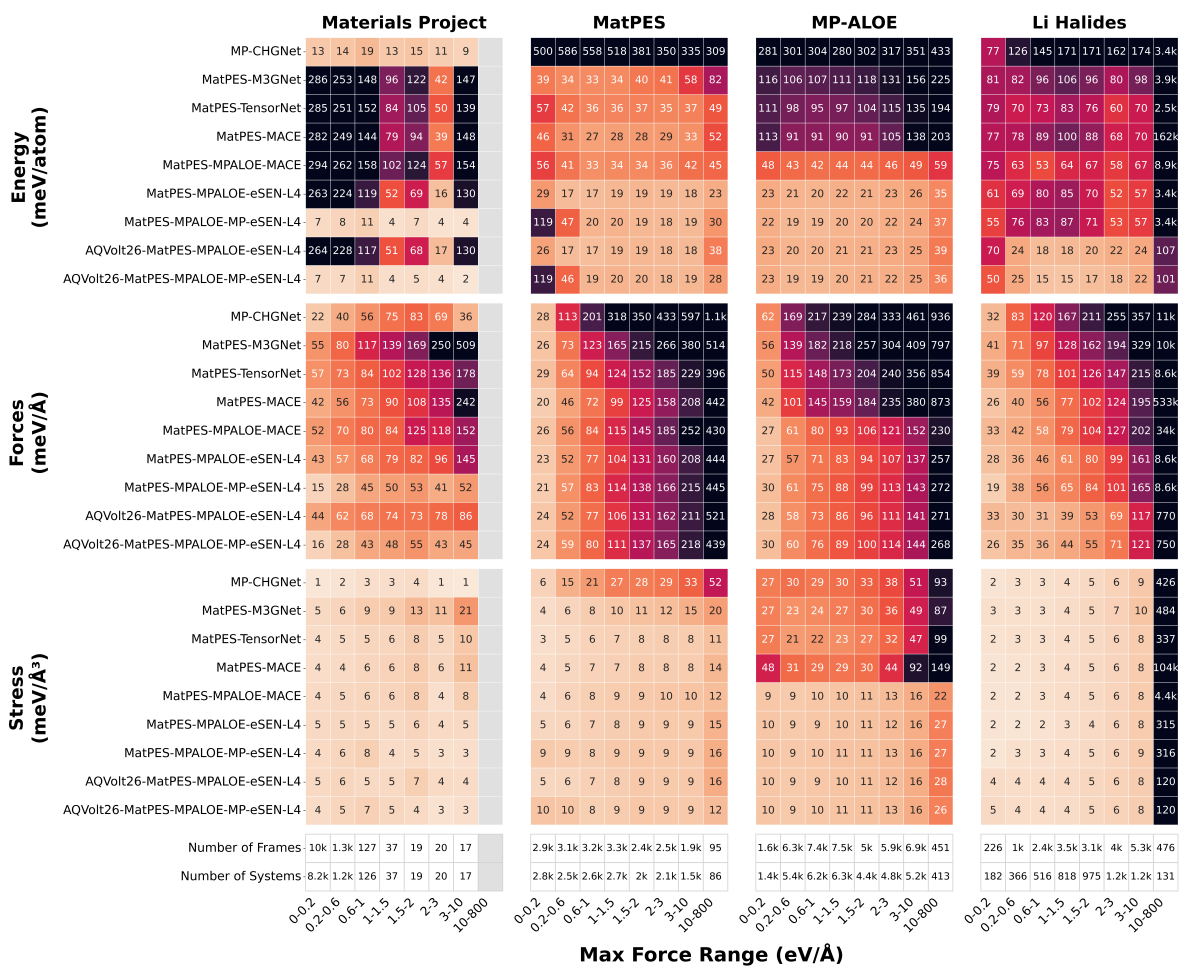}\\
    \vspace{0.5em}
    \caption{Benchmarking of energy, force, and stress predictions across trained models, evaluated on four r$^2$SCAN datasets and binned by maximum DFT force magnitude. The Materials Project \cite{Huang2025Cross-functionalPotentials}, MatPES \cite{Kaplan2025AMaterials}, and MP-ALOE \cite{Kuner2025MP-ALOE:Potentials} sets explicitly exclude Li-halide systems. The ``Li Halides'' set aggregates all halide configurations from these sources, AQVolt26, and an r$^2$SCAN-recomputed subset of GNoME \cite{Merchant2023ScalingDiscovery}, sampled from MLMD trajectories. Test splits consist of in-distribution configurations strictly unseen by the eSEN models; however, publicly available baseline checkpoints may have previously encountered this data as exact training holdouts are unavailable. Further granular analysis in the Supplementary Information reveals a minor trade-off: adapting to the highly perturbed AQVolt26 configurational space slightly degrades prediction precision on the small, highly stable GNoME subset.}
    \label{fig:benchmark_bins}
\end{figure}

Figure \ref{fig:benchmark_bins} illustrates the comparative first-principles benchmarking of various models, binned by maximum force magnitude, to reveal how predictive fidelity changes as systems move away from equilibrium. Unsurprisingly, MP-CHGNet, trained exclusively on near-equilibrium data, exhibits severe degradation across all off-equilibrium tasks. Conversely, models exposed to high-energy data (e.g., MACE, TensorNet, and eSEN) demonstrate relatively robust force and energy predictions across diverse off-equilibrium regimes. When evaluating these off-equilibrium models on the MP holdout, the energy prediction errors appear to fluctuate more significantly. Unsurprisingly, adding MP produces a dramatic improvement on near-equilibrium relaxation-like regimes. However, the same addition does not improve other tasks, where force and stress are unchanged or slightly worse, and marginally improves energy. The slight degradation in local gradients is potentially exacerbated by the use of differing pseudopotentials for several elements, largely the lanthanides, between the MP and non-MP datasets. Excluding these samples during data aggregation could potentially alleviate this conflict in future training efforts. However, the MP should be interpreted as a task-specific near-equilibrium anchor rather than as a uniformly beneficial addition across all benchmarks.

The inclusion of off-equilibrium data stabilizes predictions across standard molecular dynamics regimes. Energy, forces, and stresses are particularly more challenging to gauge by training on MatPES and MP-ALOE alone. For halogenated Li materials, the AQVolt26 models perform better notably for strained configurations with maximum force magnitudes of 0.2 eV/\AA{} and beyond with energy errors of 15-24 vs. 52-85 meV/atom on eSEN equivalents. In the most extreme force regime of AQVolt26 (10--800 eV/\AA), likely corresponding to severely overlapping atomic radii or highly strained configurations, almost all models-including those explicitly co-trained on AQVolt26 experience sharp degradation in accuracy. Among the other model frameworks, MatPES-MACE struggles the most at this task for this force range. 

We note that adapting to the highly perturbed halide structures unique to AQVolt26 marginally shifts the predictive baseline for stable configurations in GNoME (see Figure \ref{fig:gnome_aqvolt_benchmark}). To determine whether this shift impacts the identification of true ground states, we next evaluate the models' performance on 0 K structural relaxations.

\subsubsection{Balancing Ground-State Precision}

\begin{figure}[ht!]
    \centering
    % First Image
    \includegraphics[width=0.85\linewidth]{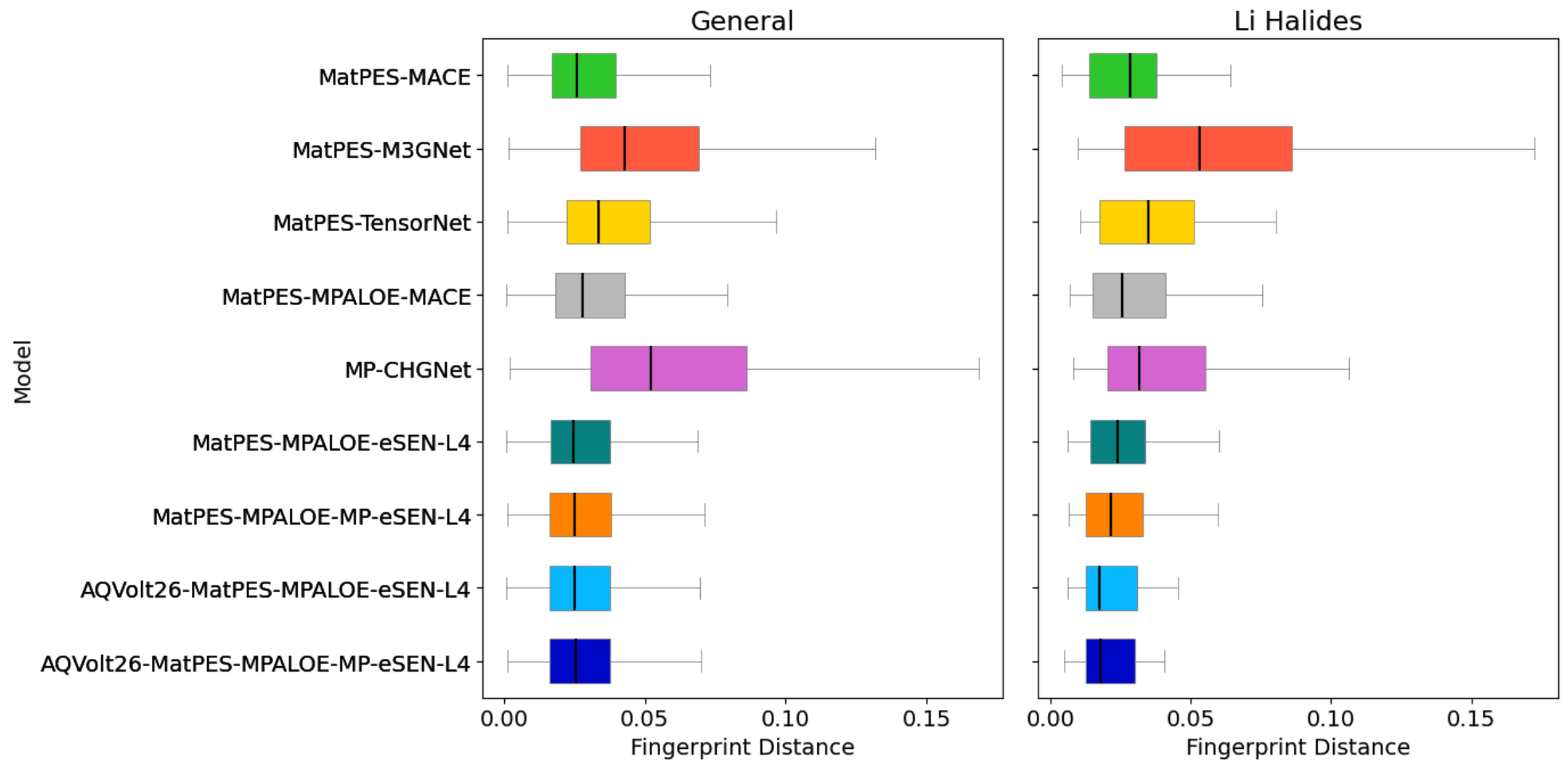}
    
    \vspace{0.3cm} % Adds a little professional breathing room between the two images
    
    % Second Image
    \includegraphics[width=0.85\linewidth]{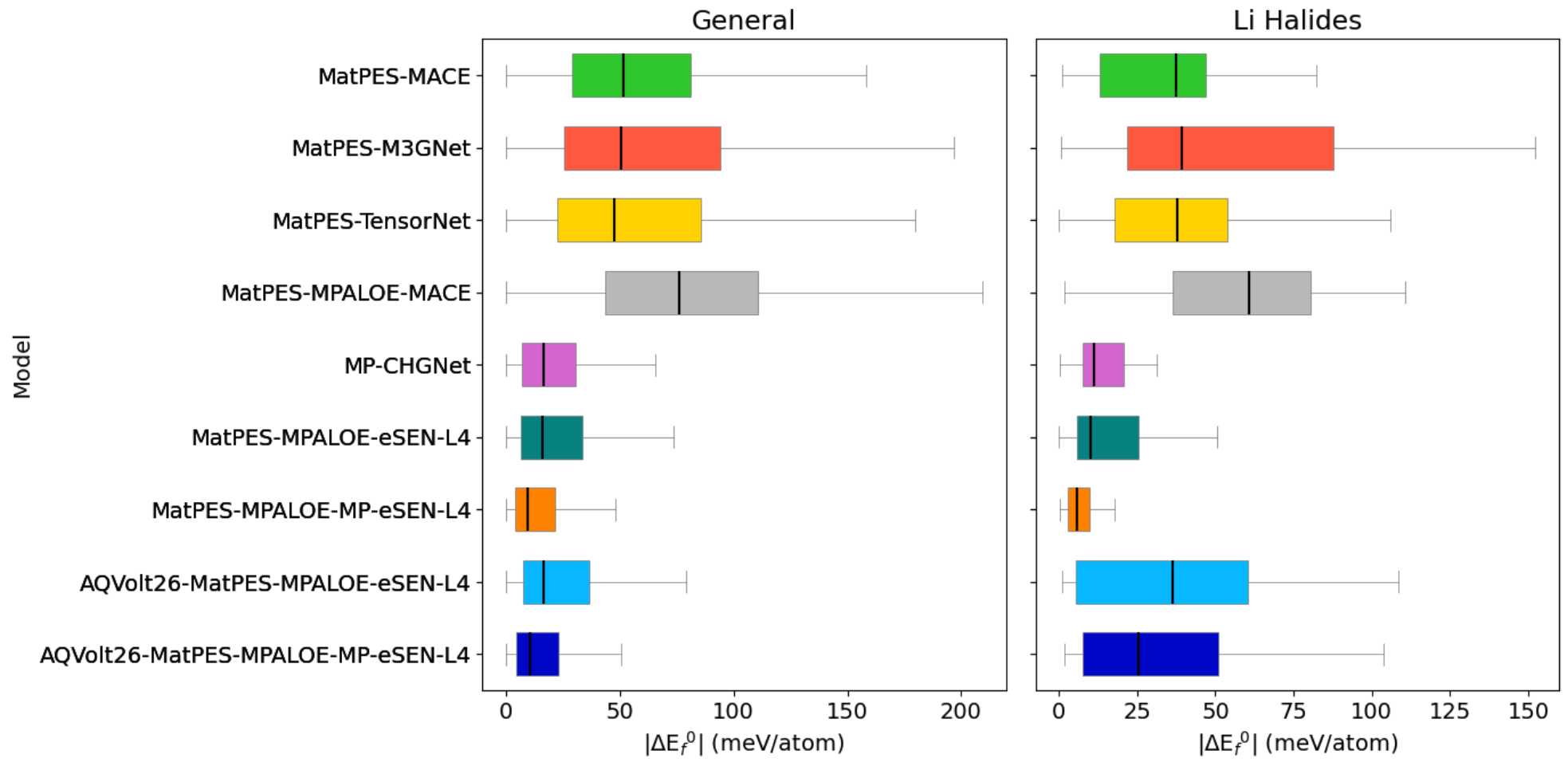}
    
    \caption{Distributions of structural similarity (top) and formation energy per atom (bottom) when comparing structures relaxed with ML interatomic potentials against r$^2$SCAN DFT relaxations. 1,000 out-of-domain structures were randomly selected from the GNoME \cite{Merchant2023ScalingDiscovery} database. Before performing geometry optimizations, every atomic site was subjected to a random perturbation of 0.1 \AA, and fingerprint distances were calculated using the CrystalNN algorithm \cite{Zimmermann2020LocalSimilarity}.}
    \label{fig:benchmark_struct_opt}
\end{figure}

While incorporating extreme off-equilibrium geometries into the training distribution improves dynamic robustness, exposing models to heavily distorted, high-temperature configurations can shift the optimization objective and degrade their precision on strict, 0 K equilibrium tasks. To evaluate whether the models suffer from this trade-off, their ability to reproduce r$^2$SCAN DFT structural relaxations was evaluated using 1,000 hold-out near-equilibrium structures (Figure \ref{fig:benchmark_struct_opt}). Fingerprint distances across all eSEN-L4 variants remain consistently low for both the general and lithium halide sets on the GNoME holdout. This indicates that exposure to high-temperature configurations does not materially degrade geometric optimization capabilities. However, distinctions emerge in formation energy predictions. On the general GNoME test set, the inclusion of AQVolt26 introduces an increase in error and variance. This effect is more pronounced within the lithium halide subset, where the AQVolt26-trained eSEN model exhibits higher formation energy errors compared to its non-AQVolt26 counterpart. Explicit co-training with near-equilibrium Materials Project data provides a modest reduction in median error and variance across these evaluations. This behavior is consistent with observations on the GNoME off-equilibrium holdouts, where adapting to distorted landscapes slightly shifted the baseline for stable configurations. This trade-off could potentially be lessened by imposing a stricter force threshold on the AQVolt26 dataset to prevent the models from over-adapting to the most highly strained geometries. We leave such ablation studies for future investigation. Nevertheless, the absolute error ranges currently achieved remain within workable margins for practical screening on GNoME.

\subsubsection{Dynamic Stability and Physical Consistency}

\begin{figure}[ht!]
    \centering
    \includegraphics[width=0.48\linewidth]{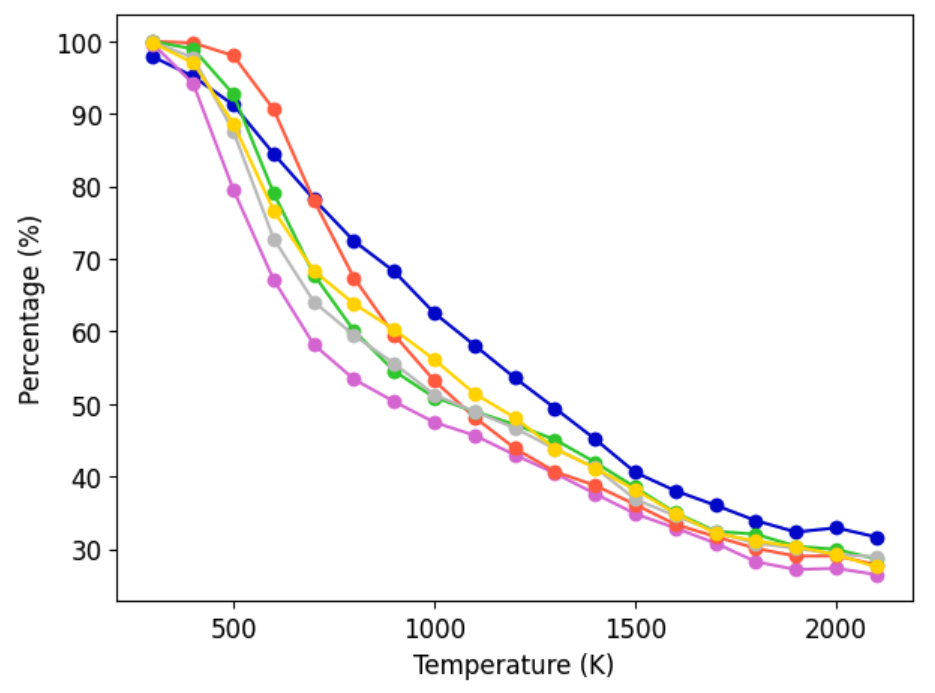}
    \includegraphics[width=0.50\linewidth]{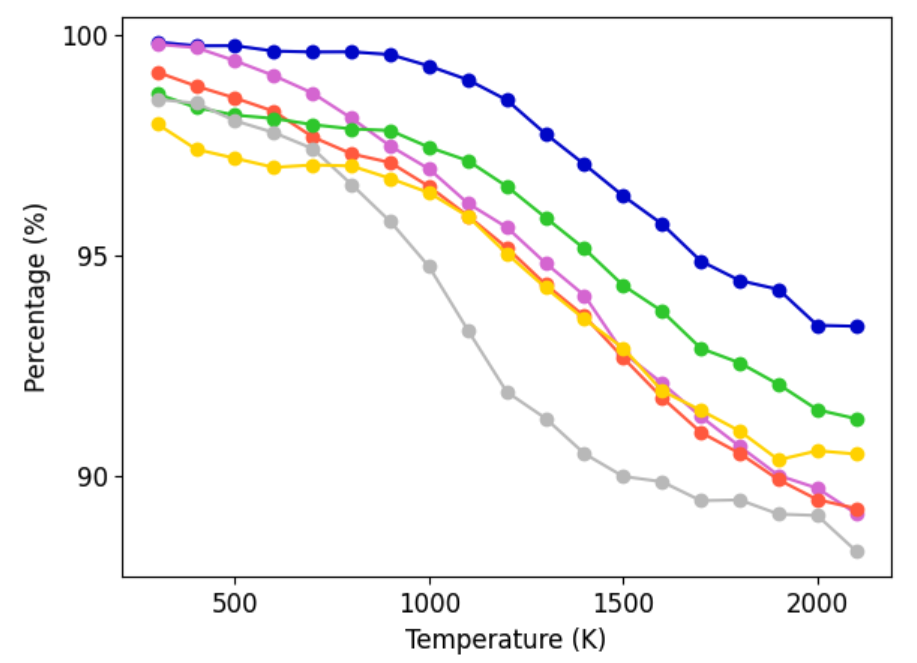}\\
    \vspace{0.5em}
    \includegraphics[width=0.7\linewidth]{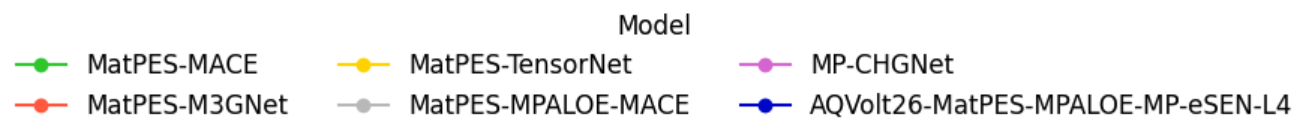}
    \caption{NpT MD controlled heating simulations (300--2,100 K) conducted across ML interatomic potentials for 50 ps with a 1 fs time step at 1 bar. Each model and material was evaluated three times on a subset of 25 Li halide (left) and 100 non-Li halide (right) out-of-domain GNoME \cite{Merchant2023ScalingDiscovery} materials. Termination steps are defined heuristically by significant volume expansion or the loss of atoms.}
    \label{fig:md_stability}
\end{figure}

Having established that the optimally trained models maintain fidelity at equilibrium, we evaluated their performance in dynamic, high-temperature environments. The primary objective of this evaluation is to determine whether the inclusion of the AQVolt26 dataset---which intentionally features highly distorted, higher cohesive energy configurations---inadvertently compromises the overall dynamic stability of the resulting potentials. 

Rigorous linear heating tests were conducted from 300 to 2,100 K (Figure \ref{fig:md_stability}). On the set of 25 Li halides (left), survival rates universally decline at a much steeper rate across all architectures. However, the AQVolt26-trained eSEN models successfully preserve the essential physical guardrails established by the MatPES and MP-ALOE baselines, exceeding in performance for high temperature environments. The eSEN models maintain robust stability on par with foundational potentials up to a notable cross-over point at approximately 700 K. Before this threshold, survival rates diverge slightly, likely due to the ample representation of more perturbed halides in AQVolt26. Despite this, the AQVolt26 models still exhibit high physical integrity ($>$80\% survival). On the GNoME halogen-free subset (right), we observe a drastically different trend: The AQVolt26-eSEN model demonstrates highly competitive performance, maintaining the highest survival rate at all temperatures. Surprisingly, MatPES-MPALOE-MACE seems to struggle relative to an equilibrium-state-trained model such as MP-CHGNet, which could be a result of the model architecture and training differences. Further benchmarks need to be conducted with the other non-AQVolt eSEN models to clarify these discrepancies, in addition to pressure ramping to distinguish model performance even further.

As an initial step toward this necessary decoupling, we can evaluate the underlying physical consistency of the models using static stress tests. By artificially straining the lattice, we isolate the model's algorithmic stability under extreme volume changes without the confounding variable of kinetic temperature. Table \ref{tab:mlip_arena} presents one such benchmark developed by MLIP Arena \cite{Chiang2025MLIPPlatform}, which evaluates the PES under extreme hydrostatic strain. This test assesses how models handle uniform lattice deformations of $\pm$20\% in all lattice parameters using single-point calculations. To pass, systems must exhibit a single global minimum in energy; monotonic increases in energy ($-1$ for compressive) and $\frac{\partial E}{\partial V}$ ($+1$ for tensile) should be noted with deformation based on the Spearman's coefficient.

\begin{table}[H]
\centering
\captionsetup{width=0.9\linewidth}
\caption{Evaluating the potential energy surface under extreme hydrostatic strain with an Energy-Volume scan benchmark on 1,000 structures developed by MLIP Arena. $^*$Results for baseline models were derived from the MP-ALOE study \cite{Kuner2025MP-ALOE:Potentials}.}
\label{tab:mlip_arena}
\begin{tabular}{lccc}
\toprule
Model & Failures (\%) & \multicolumn{2}{c}{Spearman's coefficient} \\
&  & $E$ & $\frac{\partial E}{\partial V}$ \\
\midrule
MatPES-MACE$^*$ & 14.8 & -0.919 & 0.857 \\
MatPES-M3GNet & 1.1 & -1.000 & 0.993 \\
MatPES-TensorNet & 7.7 & -0.997 & 0.894 \\
MatPES-MPALOE-MACE$^*$ & 2.5 & -0.990 & 0.980 \\
MP-CHGNet & 0.0 & -1.000 & 0.962 \\
% \bf{AQVolt26-eSEN-L4} & \bf{0.1} & \bf{-1.000} & \bf{1.000} \\
% AQVolt26-MP-eSEN-L4-magmom & 16.9 & -0.947 & 0.907 \\ 
% AQVolt0.25-MP1.00-eSEN-L4 & 9.1 & -0.982 & 0.923 \\ 
MatPES-MPALOE-eSEN-L4 & 0.1 & -1.00 & 0.996 \\
MatPES-MPALOE-MP-eSEN-L4 & 13.0 & -0.96 & 0.929 \\
AQVolt26-MatPES-MPALOE--eSEN-L4 & 0.2 & -1.00 & 1.000 \\
AQVolt26-MatPES-MPALOE-MP-eSEN-L4 & 12.9 & -0.968 & 0.927 \\
% MatPES-MP-ALOE-MP-AQVolt-eSEN-L4-V2 & 12.8 & -0.967 & 0.926 \\ 
\bottomrule
\end{tabular}
\end{table}

While future work should establish a broader suite of decoupled thermodynamic benchmarks for soft materials, this static E-V scan confirms that the AQVolt26-MatPES-MPALOE-eSEN-L4 model possesses exceptional stability. It exhibits only a 0.2\% failure rate when subjected to $\pm$20\% uniform lattice deformations, outperforming several state-of-the-art baselines. The models trained on off-equilibrium data predict stiffer, more rigid bonds based on the higher bulk moduli obtained from the energy-volume curvatures, which may suggest that training on these datasets may help capture the steep changes in the PES for perturbed structures (Figure \ref{fig:mlip_arena}).

% This strongly suggests that the high-temperature volume expansions observed during the halide MD simulations are driven by genuine physical lattice softening rather than a catastrophic breakdown of the model's predicted potential energy surface.

Ultimately, this robustness enables the accurate prediction of downstream battery kinetics, such as ionic conductivity. To further probe this, MLMD simulations were leveraged to evaluate lithium-based solid-state electrolytes, benchmarking the results against experimental data from the literature \cite{Therrien2025OBELiX:Electrolytes}. The AQVolt26-trained model demonstrates enhanced predictive accuracy for room-temperature conductivity for lithium halide systems with a MAE of 0.6 vs. 4.2 mS/cm relative to MatPES-TensorNet (Table \ref{tab:ionic_conductivity_lihalides}), notably achieving these gains without inducing catastrophic failures in non-halide chemistries. Over-estimating the conductivity is observed nearly twice as often with MatPES-TensorNet for Li halides, which is prevalent with ML force-fields relative to DFT due to the softening of the PES \cite{Deng2025SystematicPotentials}. The AQVolt26-trained model appears to be more conservative with these mobility predictions, which reduces false positives and allows for careful investigation of truly promising candidates. Although both models were less accurate for halogen-free chemistries, overall improvements were observed with AQVolt26 an error of 149 vs. 2,081 mS/cm (Table \ref{tab:ionic_conductivity_general}). However, MatPES-TensorNet is more accurate in a little over half of these systems, which suggests that the AQVolt26-trained model may have lost some chemical intuition that may be especially valuable for ion transport for general compositions. Although MatPES-TensorNet was selected due to its comparable performance in the MD stability benchmark for halogenated systems, a more thorough investigation would require contrasting the AQVolt26-eSEN-L4 model with a MatPES-eSEN equivalent to isolate the improvements of the dataset from the differences in model architecture.

%%%%%%%%%%%%%%%%%%%%%%%%%%%%%%%%%%%%%%%%%%%%%%%%%%%%%%%%%%%%%%%%%%%%%%%%%%%%%%%%%%%%%%%%%%

\section{Discussion}

The accelerated discovery of solid-state electrolytes relies on atomistic simulations that are both computationally efficient and physically robust at elevated temperatures. In this work, we introduced AQVolt26, a specialized dataset of 322,656 r$^2$SCAN calculations explicitly designed to map the highly anharmonic, molten-sublattice configurational landscape of lithium halide materials. The central result of this study is that targeted halide configurational sampling, rather than broader compositional coverage alone, is required to obtain accurate r$^2$SCAN ML interatomic potentials for high-temperature lithium-halide dynamics. Using matched eSEN models trained with and without AQVolt26, we show that AQVolt26 produces an order-of-magnitude improvement on the intended off-equilibrium halide regime while preserving competitive performance on established foundational benchmarks.

At the same time, the present benchmarks show that the Materials Project (MP) relaxation data should be viewed as a specialized complement rather than a universally beneficial addition. Its data strongly improves near-equilibrium relaxation benchmarks, but it does not support other regimes in force accuracy, only modestly increases AQVolt26 energy accuracy, and substantially worsens extreme-strain physical-consistency metrics. Accordingly, two use cases emerge from this work. For high-temperature halide molecular dynamics, PES robustness under strong distortion, and transport-oriented screening, AQVolt26-MatPES-MPALOE-eSEN-L4 is the more physically consistent default. For relaxation-heavy workflows and near-equilibrium property prediction, AQVolt26-MatPES-MPALOE-MP-eSEN-L4 is the preferred model. 

More broadly, we reinforce a point also emphasized by MatPES and MP-ALOE, that dataset composition is the primary determinant of where a universal potential succeeds. MatPES shows that carefully sampled single-point data can outperform much larger but less targeted datasets, while MP-ALOE shows that broader pressure and off-equilibrium coverage can improve MD stability and E–V behavior when combined with MatPES. AQVolt26 utilizes halide solid-state electrolytes as a targeted testing ground for foundational models. Our results demonstrate that while universal r$^2$SCAN baselines provide a robust foundation, modeling dynamically soft materials benefits significantly from targeted high-temperature data. Because soft materials exhibit shallow anharmonic potential energy surfaces, explicit exposure to highly distorted phase spaces improves dynamic reliability. We anticipate that this domain-specific sampling strategy is conceptually transferable. Applying this methodology to structurally analogous solid-state chemistries, such as sulfides and complex hydrides \cite{Lee2025Multi-Solid-ElectrolyteProspects}, could provide a practical pathway to advance machine learning interatomic potentials for next-generation battery research.

%%%%%%%%%%%%%%%%%%%%%%%%%%%%%%%%%%%%%%%%%%%%%%%%%%%%%%%%%%%%%%%%%%%%%%%%%%%%%%%%%%%%%%%%%%

\section*{Author Contributions}

\textbf{J.K.} performed ML model training, executed the data generation and surrogate-driven sampling workflows, generated benchmark datasets, analyzed the data, and co-wrote the manuscript. \\
\textbf{C.W.} advised on data generation strategies. \\
\textbf{A.X.} co-conceived the initial project concept and provided strategic guidance. \\
\textbf{A.R.S.} advised on the framing of the study and the conceptual clarity of the manuscript's narrative. \\
\textbf{P.A., A.N.} assisted with DGX H100 computing infrastructure.
\textbf{C.W., T.S., S.A.} provided insightful discussions, conceptual feedback, and reviewed the manuscript. \\
\textbf{O.A.} co-conceived the project, supervised the research, performed ML model training, conducted ablation studies, analyzed the data, and co-wrote the manuscript.

\section{Acknowledgments}
The authors appreciate the support provided by the SandboxAQ leadership and business development team, notably Adam Lewis, Nadia Harhen, Arman Zaribafiyan, Jeff Graf, Scott Healey, JP Murgueitio, Takeshi Yamazaki, Jessica Pan, Stefan Leichenauer, Andrew McLaughlin, and Jack Hidary.

% [todo] acknowledge Eric Wang from Samsung?
\clearpage
\bibliographystyle{unsrtnat} 
\bibliography{references}
%%%%%%%%%%%%%%%%%%%%%%%%%%%%%%%%%%%%%%%%%%%%%%%%%%%%%%%%%%%%%%%%%%%%%%%%%%%%%%%%%%%%%%%%%%

\clearpage

% 1. Create a clear header for the SI
\begin{center}
\LARGE \textbf{Supporting Information:}\\
\vspace{0.5cm}
\Large \textbf{AQVolt26: A High-Temperature r$^2$SCAN Halide Dataset and Universal Potentials for Solid-State Batteries}
\end{center}
\vspace{1cm}

% 2. Reset counters and add the "S" prefix
\setcounter{table}{0}
\renewcommand{\thetable}{S\arabic{table}}
\setcounter{figure}{0}
\renewcommand{\thefigure}{S\arabic{figure}}
\setcounter{equation}{0}
\renewcommand{\theequation}{S\arabic{equation}}
\setcounter{section}{0}
\renewcommand{\thesection}{S\arabic{section}}
\setcounter{page}{1}
\renewcommand{\thepage}{S\arabic{page}}

\begin{table}[hbt!]
\centering
\caption{Hyperparameters for the two-stage training of the equivariant Smooth Energy Network (eSEN) force field. Stage 1 consists of direct-force pre-training, and Stage 2 consists of conservative force fine-tuning.}
\label{tab:hyperparameters}
\begin{tabular}{lcc}
\toprule
\textbf{Parameter} & \textbf{Stage 1: Pre-training} & \textbf{Stage 2: Fine-tuning} \\
\midrule
\multicolumn{3}{l}{\textit{Architecture Details}} \\
Number of Interaction Layers & \multicolumn{2}{c}{4} \\
Sphere Channels & \multicolumn{2}{c}{128} \\
Edge / Hidden Channels & \multicolumn{2}{c}{128} \\
Maximum Spherical Degree ($l_{max}$) & \multicolumn{2}{c}{2} \\
Maximum Spherical Order ($m_{max}$) & \multicolumn{2}{c}{2} \\
Radial Basis Function & \multicolumn{2}{c}{Gaussian} \\
Number of Distance Basis & \multicolumn{2}{c}{64} \\
Cutoff Radius & \multicolumn{2}{c}{6.0 \AA} \\
\midrule
\multicolumn{3}{l}{\textit{Graph Construction and Targets}} \\
Maximum Neighbors & 30 & 300 \\
Direct Force Prediction & True & False (Conservative) \\
Regress Stress & False & True \\
Precision & bfloat16 & FP32 \\
\midrule
\multicolumn{3}{l}{\textit{Optimization Schedule}} \\
Optimizer & \multicolumn{2}{c}{AdamW} \\
Global Batch Size (atoms) & 80,000 & 48,000 \\ % Added Batch Size
H100 Cards & 8 & 32 \\
%Training Steps & 10,000 & 20,000 \\
Peak Learning Rate & $8 \times 10^{-4}$ & $4 \times 10^{-4}$ \\
Learning Rate Scheduler & \multicolumn{2}{c}{Cosine Annealing} \\
Warmup Factor & 0.2 & 0.2 \\
Warmup Fraction & 0.02 & 0.01 \\
Minimum Learning Rate Factor & 0.05 & 0.01 \\
Weight Decay & $1 \times 10^{-3}$ & $1 \times 10^{-3}$ \\
Gradient Clipping Norm & 100 & 100 \\
\midrule
\multicolumn{3}{l}{\textit{Loss Coefficients}} \\
Energy Coefficient & 10 & 20 \\
Force Coefficient & 30 & 2 \\
Stress Coefficient & 0 & 1 \\
\bottomrule
\end{tabular}
\end{table}

\begin{figure}[ht!]
    \centering
    \includegraphics[scale=0.25]{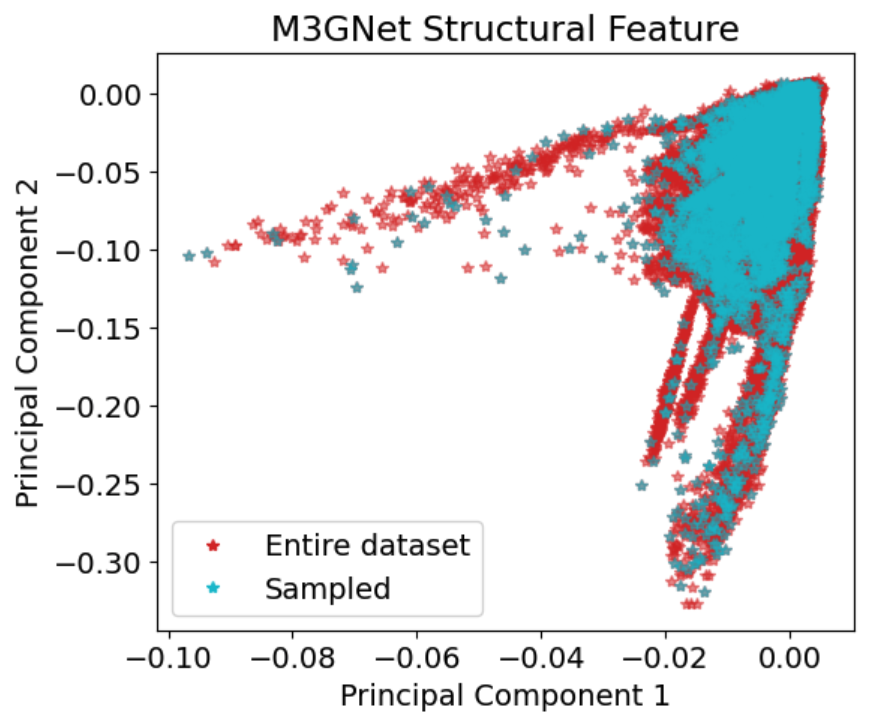}
    \includegraphics[scale=0.25]{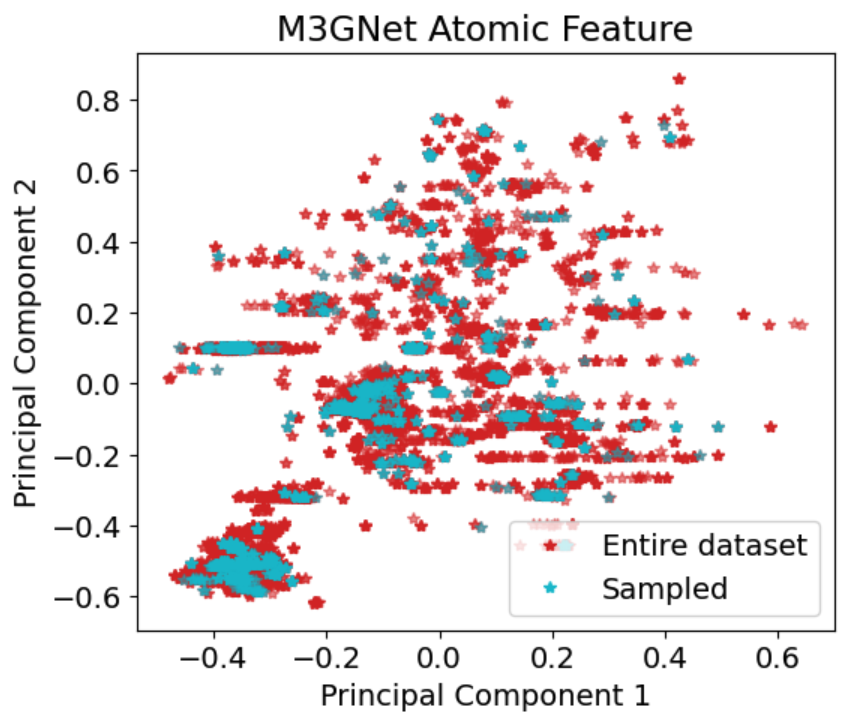}
    \caption{2DIRECT sampling \cite{Qi2024RobustSampling} after ML molecular dynamics simulations on GNoME \cite{Merchant2023ScalingDiscovery} Li halides to produce the holdout test dataset.}
    \label{fig:gnome_holdout_sampling}
\end{figure}

\begin{figure}[ht!]
    \centering
    \includegraphics[scale=0.175]{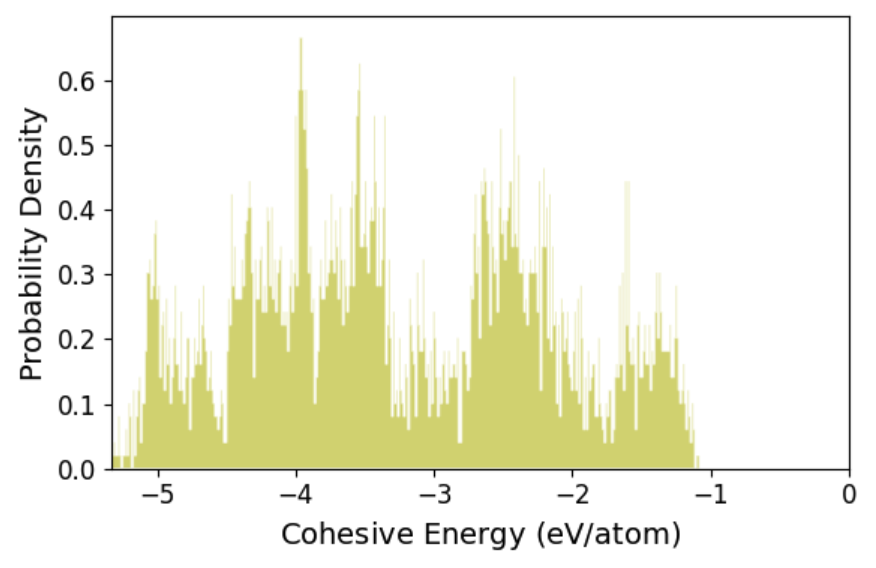}
    \includegraphics[scale=0.175]{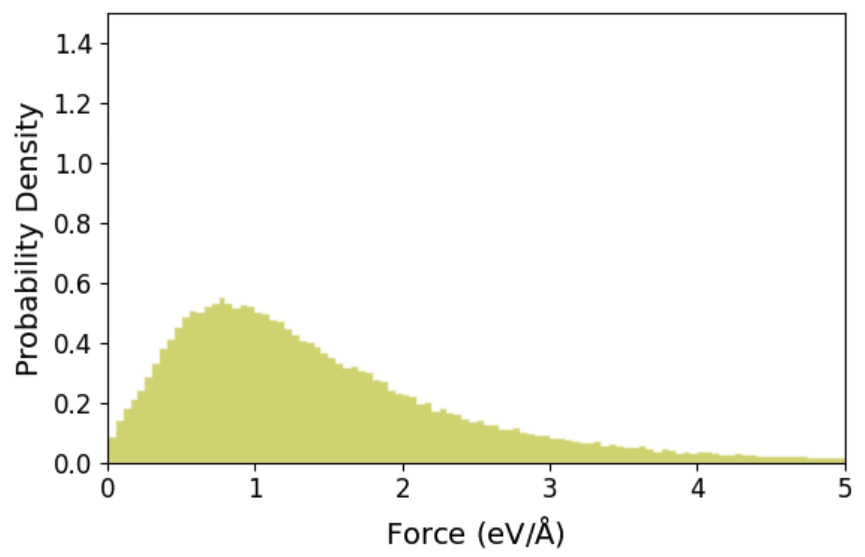}
    \includegraphics[scale=0.175]{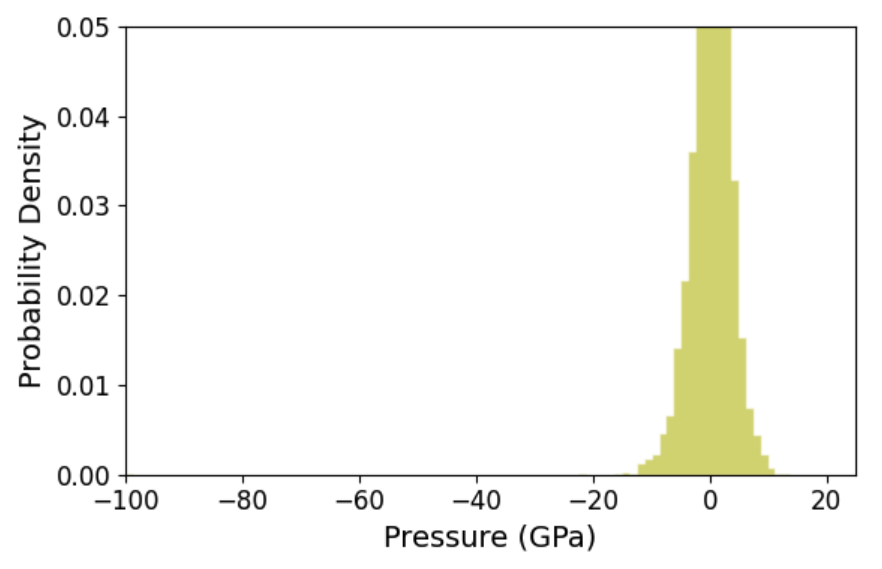}
    \caption{Distributions of cohesive energies, interatomic force magnitudes, and pressures for the GNoME holdout set generated for Li halides.}
    \label{fig:distribution_gnome}
\end{figure}

\begin{figure}[ht!]
    \centering
    \includegraphics[scale=0.3]{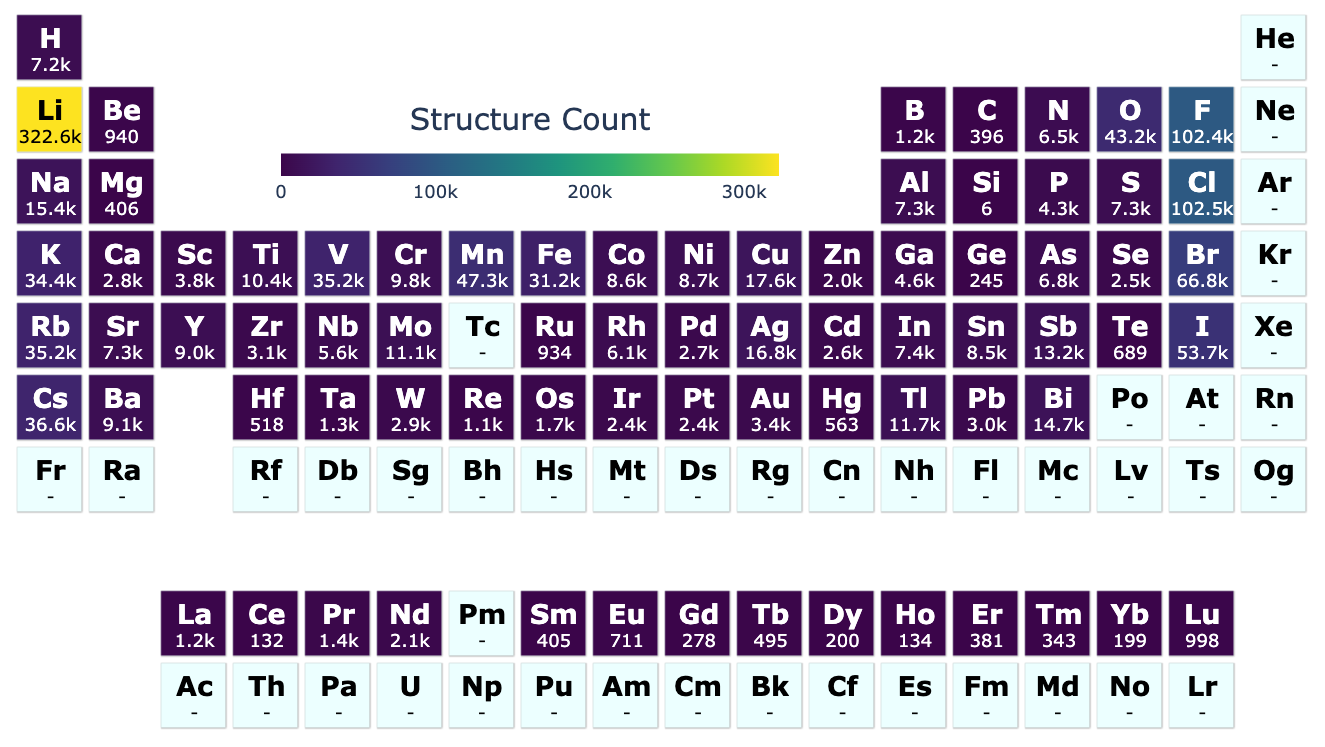}
    \includegraphics[scale=0.3]{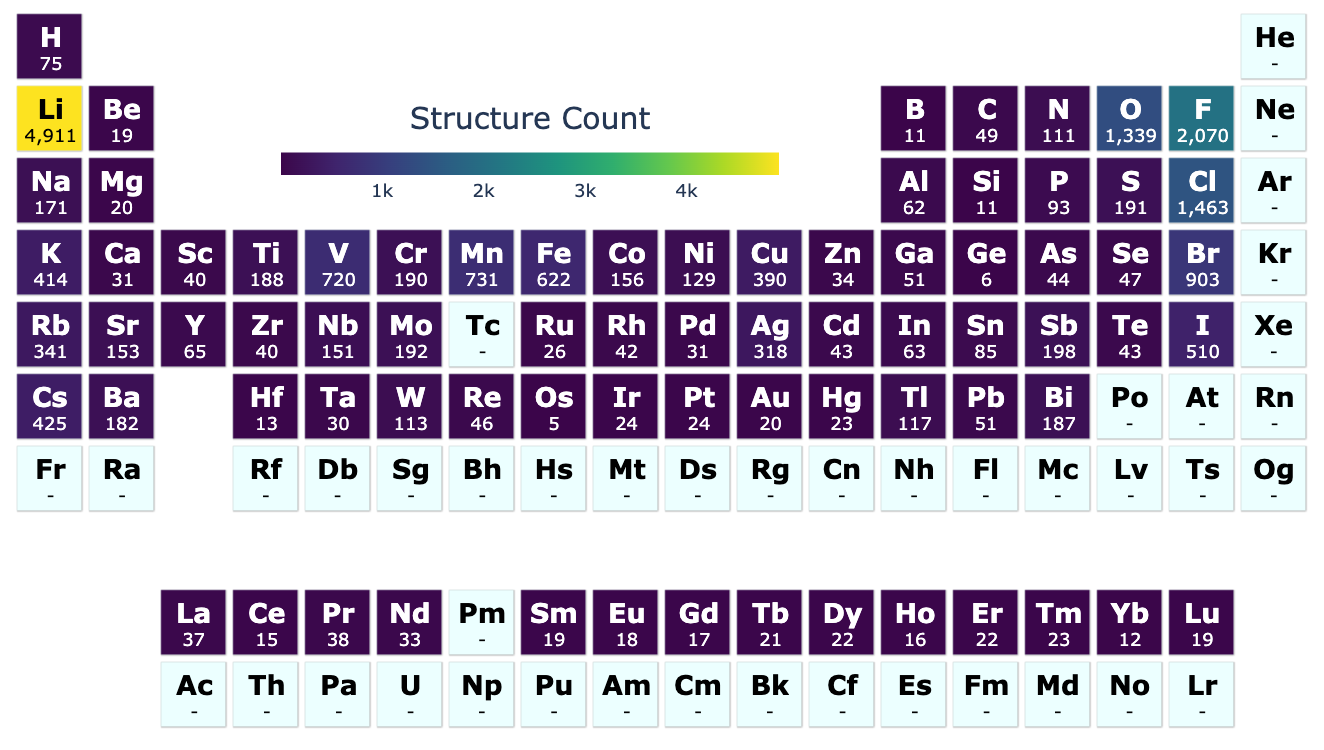}
    \caption{AQVolt26 r$^2$SCAN dataset (top) and systems (bottom) elemental distribution, plotted with pymatviz \cite{Riebesell2022Informatics}. Counts represent the number of materials with the given element.}
    \label{fig:periodic_table}
\end{figure}

\begin{figure}[ht!]
    \centering
    \includegraphics[scale=0.162]{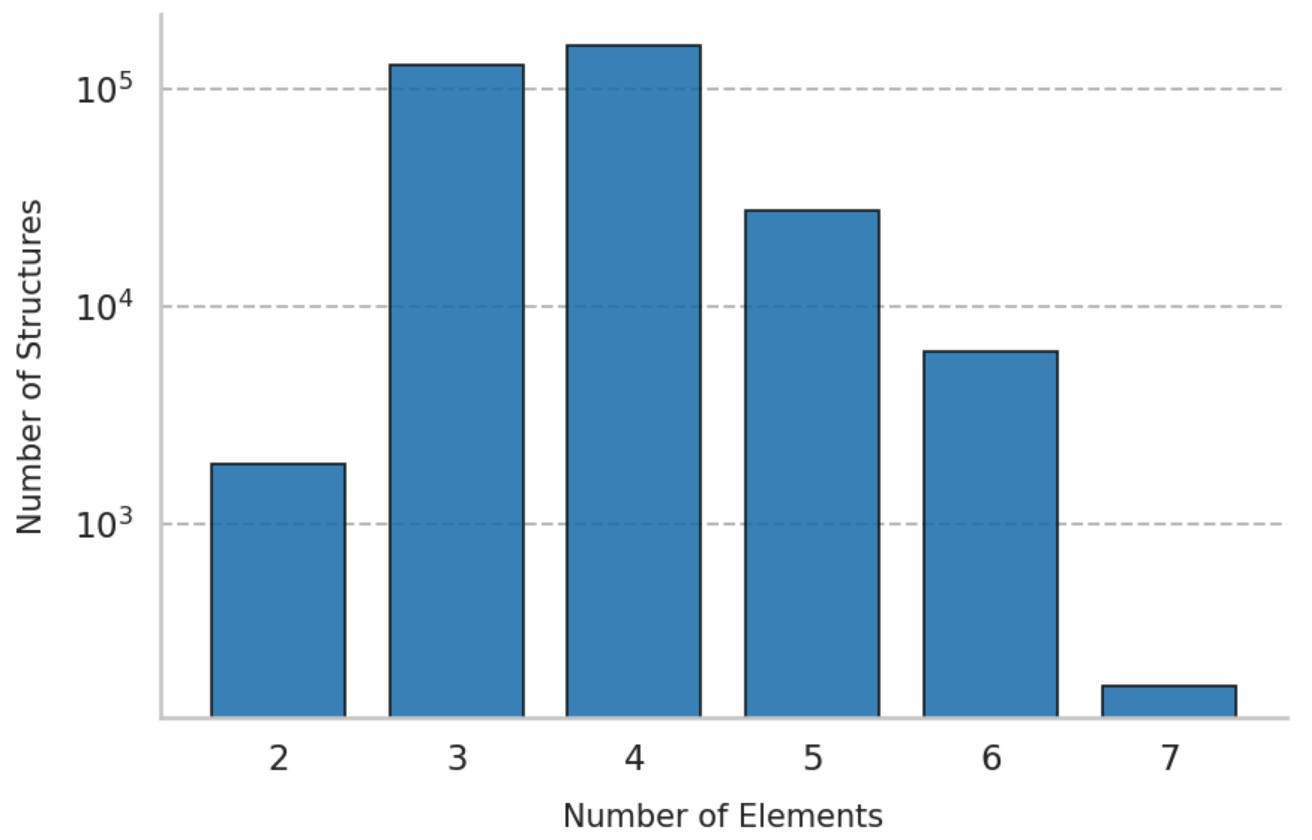}
    \includegraphics[scale=0.162]{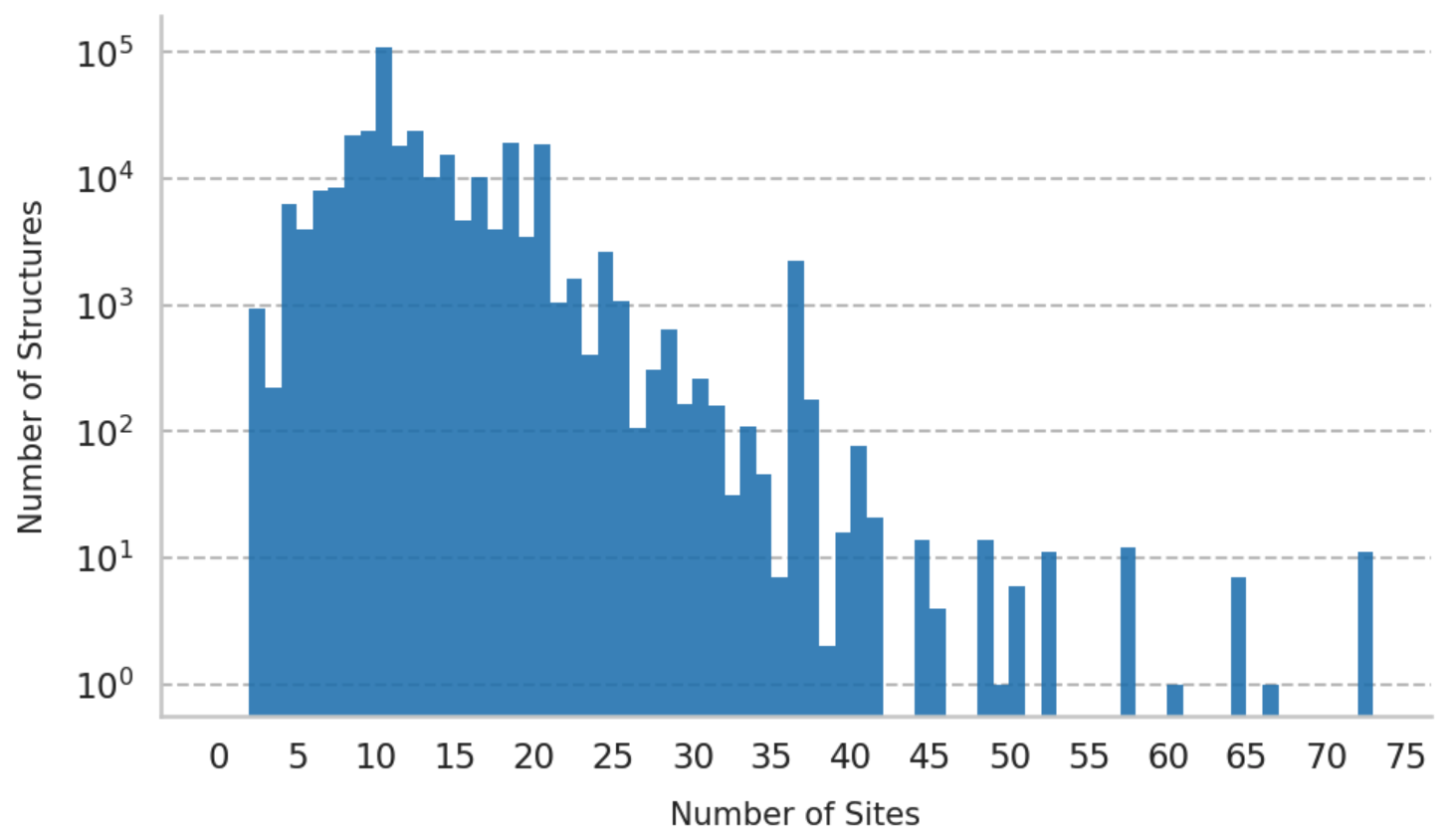}
    \caption{AQVolt26 r$^2$SCAN structure distribution by element (left) and site count (right). Structure frequency is shown on a logarithmic scale for improved visibility.}
    \label{fig:element_site_distribution}
\end{figure}

\begin{table}[htbp]
    \centering
    \caption{Mean and standard deviation of cohesive energy, force magnitudes, and pressure across evaluated datasets.}
    \label{tab:dataset_metrics}
    \renewcommand{\arraystretch}{1.2} % Adds a little padding between rows
    \begin{tabular}{lccc}
        \toprule
        \textbf{Dataset} & \textbf{Cohesive Energy} (eV/atom) & \textbf{Forces} (eV/\AA) & \textbf{Pressure} (GPa) \\
        \midrule
        Materials Project & $-4.12 \pm 0.79$ & $0.04 \pm 0.08$ & $0.03 \pm 0.52$  \\
        MatPES & $-3.91 \pm 0.98$ & $0.91 \pm 3.88$ & $-3.08 \pm 27.89$ \\
        MP-ALOE & $-2.99 \pm 1.13$ & $0.52 \pm 1.28$ & $-2.98 \pm 14.56$ \\
        AQVolt26 & $-3.33 \pm 0.76$ & $1.25 \pm 5.27$ & $-1.72 \pm 23.37$ \\
        % GNoME & $-3.26 \pm 1.10$ & $1.57 \pm 1.55$ & $0.26 \pm 3.13$ \\
        \bottomrule
    \end{tabular}
\end{table}

\begin{figure}[ht!]
    \centering
    \includegraphics[scale=0.355]{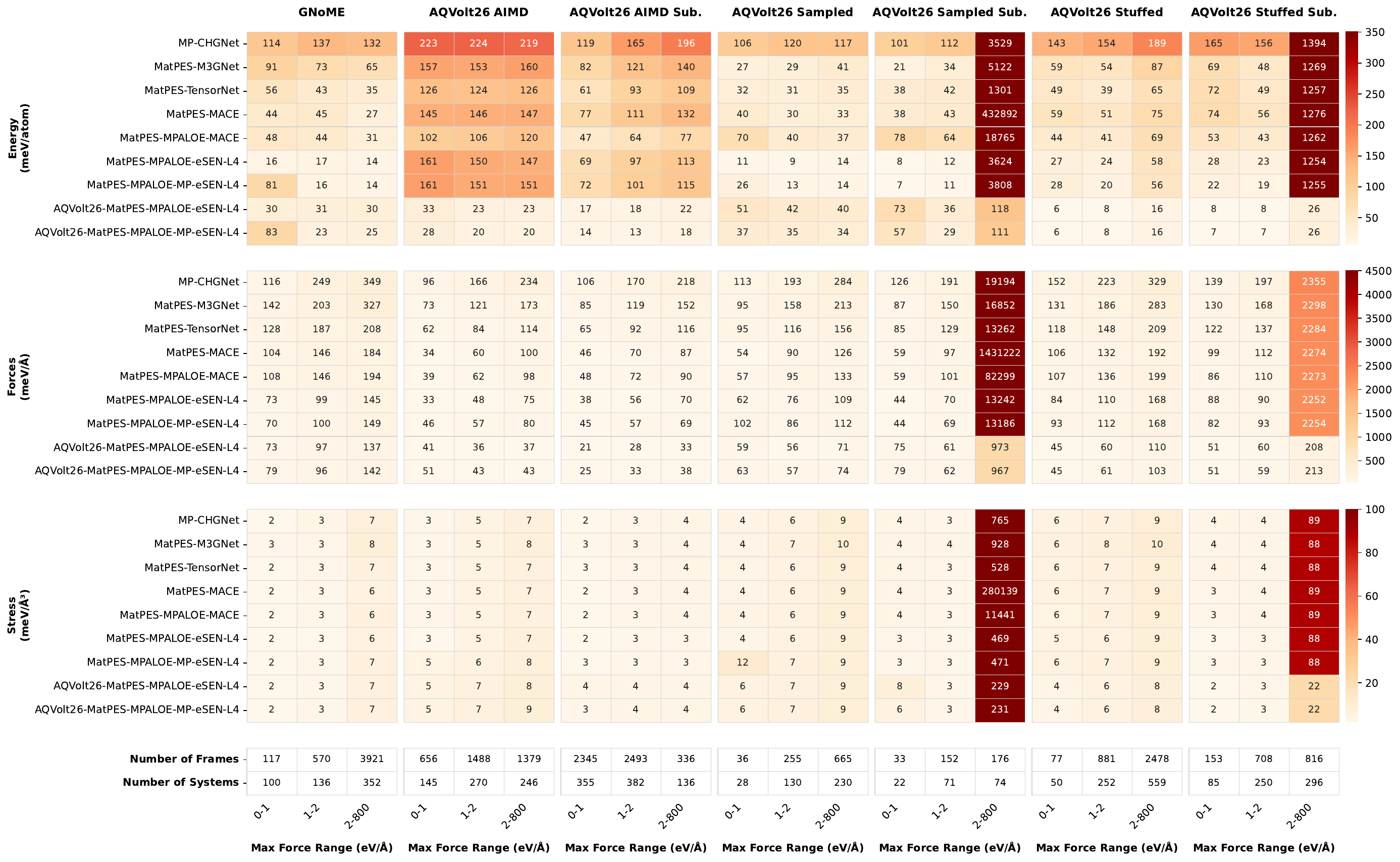}
    \caption{Benchmarking of energy, force, and stress predictions across trained models, evaluated on four r$^2$SCAN datasets and binned by maximum DFT force magnitude. This represents Li halide results for an r$^2$SCAN-recomputed subset of GNoME \cite{Merchant2023ScalingDiscovery}, sampled from MLMD trajectories, and the AQVolt26 test set. ``AIMD" without substitutions only has configurations directly from MatPES or the Materials Project, ``Sampled" represents those selected from MLMD trajectories, ``Sub." denotes structures with a halide anion substitution, and ``Stuffed" consists of lithiated empty host structures (see Methods for more details). The number of single-point frames as well as unique systems are labeled at the bottom.}
    \label{fig:gnome_aqvolt_benchmark}
\end{figure}

% \begin{figure}[ht!]
%     \centering
%     \includegraphics[scale=0.22]{pca_datasets_struct.png}
%     \includegraphics[scale=0.225]{pca_datasets_atom.png}
%     \includegraphics[scale=0.2]{pca_datasets_legend.png}
%     \caption{Principal component analysis of compositional space of Li halide materials in the MatPES, GNoME, AQVolt26 datasets.}
%     \label{fig:pca_comparison}
% \end{figure}

\begin{figure}[ht]
    \centering
    \includegraphics[width=0.7\linewidth]{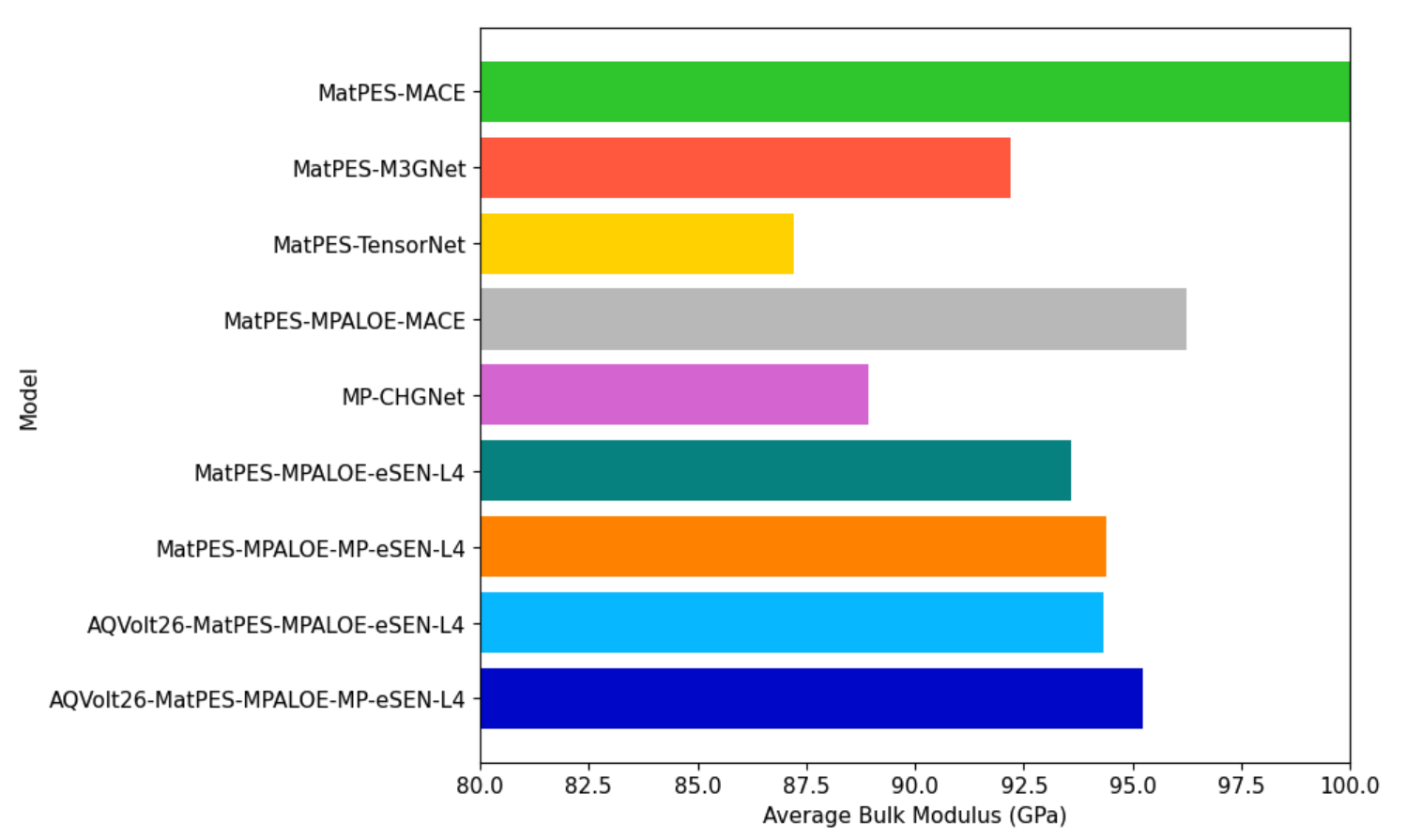}
    \caption{Bulk moduli predicted from analyzing the curvatures of energy vs. volume changes from the MLIP-Arena benchmark. The plot is zoomed in and prematurely cuts off the 577 GPa result for MatPES-MACE.}
    \label{fig:mlip_arena}
\end{figure}

\clearpage

\begin{table}[ht!]
\centering
\small
\caption{Ionic conductivity benchmarking on 9 Li halogen-containing solid-state electrolytes relative to experimental values, with reference to experimentally-measured values from the OBELiX dataset \cite{Therrien2025OBELiX:Electrolytes} using the MatPES-TensorNet and AQVolt26-MatPES-MPALOE-MP-eSEN-L4 models.}
\label{tab:ionic_conductivity_lihalides}
\begin{tabular}{lcccc}
\hline
& \multicolumn{2}{c}{\textbf{MatPES-TensorNet}} & \multicolumn{2}{c}{\textbf{AQVolt26-eSEN-L4}} \\
\textbf{Composition} & \textbf{Ionic Conductivity (mS/cm)} & \textbf{Error (mS/cm)} & \textbf{Ionic Conductivity (mS/cm)} & \textbf{Error (mS/cm)} \\ \hline
$\text{Li}_{14}\text{Nd}_{5}(\text{Si}_{11}\text{N}_{19}\text{O}_{5})\text{O}_{2}\text{F}_{2}$ & $1.496 \times 10^{1}$ & $1.496 \times 10^{1}$ & $9.180 \times 10^{-3}$ & $9.180 \times 10^{-3}$ \\
$\text{LiAlCl}_{4}$ & $1.106 \times 10^{1}$ & $1.106 \times 10^{1}$ & $1.245 \times 10^{-22}$ & $-1.000 \times 10^{-3}$ \\
$\text{Li}_{2}\text{CrCl}_{4}$ & $5.865 \times 10^{0}$ & $5.865 \times 10^{0}$ & $4.087 \times 10^{0}$ & $4.087 \times 10^{0}$ \\
$\text{LiGaBr}_{4}$ & $3.470 \times 10^{0}$ & $3.463 \times 10^{0}$ & $3.088 \times 10^{-7}$ & $-7.000 \times 10^{-3}$ \\
$\text{Li}_{6}\text{SiO}_{4}\text{Cl}_{2}$ & $2.582 \times 10^{0}$ & $2.582 \times 10^{0}$ & $1.308 \times 10^{0}$ & $1.308 \times 10^{0}$ \\
$\text{LiV}(\text{PO}_{4})\text{F}$ & $3.800 \times 10^{-2}$ & $3.719 \times 10^{-2}$ & $6.398 \times 10^{-7}$ & $-8.094 \times 10^{-4}$ \\
$\text{LiB}_{6}\text{O}_{9}\text{F}$ & $1.688 \times 10^{-2}$ & $1.688 \times 10^{-2}$ & $5.809 \times 10^{-10}$ & $5.799 \times 10^{-10}$ \\
$\text{Li}_{3}\text{OBr}$ & $4.848 \times 10^{-9}$ & $-1.100 \times 10^{-3}$ & $8.030 \times 10^{-15}$ & $-1.100 \times 10^{-3}$ \\
$\text{Li}_{9}\text{Mg}_{3}(\text{PO}_{4})_{4}\text{F}_{3}$ & $1.437 \times 10^{-11}$ & $1.337 \times 10^{-11}$ & $7.784 \times 10^{-8}$ & $7.783 \times 10^{-8}$ \\ \hline
\end{tabular}
\end{table}

\begin{table}[ht!]
\centering
\scriptsize
\caption{Ionic conductivity benchmarking on 63 Li halogen-free solid-state electrolytes relative to experimental values, with reference to experimentally-measured values from the OBELiX dataset \cite{Therrien2025OBELiX:Electrolytes} using the MatPES-TensorNet and AQVolt26-MatPES-MPALOE-MP-eSEN-L4 models.}
\label{tab:ionic_conductivity_general}
\begin{tabular}{lcccc}
\hline
& \multicolumn{2}{c}{\textbf{MatPES-TensorNet}} & \multicolumn{2}{c}{\textbf{AQVolt26-eSEN-L4}} \\
\textbf{Composition} & \textbf{Ionic Conductivity (mS/cm)} & \textbf{Error (mS/cm)} & \textbf{Ionic Conductivity (mS/cm)} & \textbf{Error (mS/cm)} \\ \hline
$\text{Li}_{5}\text{AlO}_{4}$ & $1.143 \times 10^{5}$ & $1.143 \times 10^{5}$ & $1.716 \times 10^{-1}$ & $1.716 \times 10^{-1}$ \\
$\text{LiAlGeO}_{4}$ & $1.669 \times 10^{4}$ & $1.669 \times 10^{4}$ & $1.216 \times 10^{-9}$ & $1.215 \times 10^{-9}$ \\
$(\text{Li}_{0.5}\text{Ce}_{0.25}\text{Sm}_{0.25})(\text{MoO}_{4})$ & $8.660 \times 10^{-12}$ & $-1.800 \times 10^{-7}$ & $5.747 \times 10^{3}$ & $5.747 \times 10^{3}$ \\
$(\text{Li}_{0.5}\text{Ce}_{0.25}\text{Pr}_{0.25})(\text{MoO}_{4})$ & $1.218 \times 10^{-11}$ & $-1.000 \times 10^{-6}$ & $1.924 \times 10^{3}$ & $1.924 \times 10^{3}$ \\
$\text{Li}_{4}\text{P}_{2}\text{O}_{7}$ & $1.213 \times 10^{-19}$ & $-1.000 \times 10^{-12}$ & $6.235 \times 10^{2}$ & $6.235 \times 10^{2}$ \\
$\text{LiTi}_{2}(\text{PO}_{4})_{3}$ & $1.363 \times 10^{-4}$ & $-7.186 \times 10^{-2}$ & $4.208 \times 10^{2}$ & $4.207 \times 10^{2}$ \\
$\text{Li}_{3}\text{BP}_{2}\text{O}_{8}$ & $1.494 \times 10^{-28}$ & $-9.600 \times 10^{-9}$ & $2.665 \times 10^{2}$ & $2.665 \times 10^{2}$ \\
$\text{Li}_{10}\text{SnP}_{2}\text{S}_{12}$ & $5.057 \times 10^{1}$ & $4.357 \times 10^{1}$ & $6.256 \times 10^{1}$ & $5.556 \times 10^{1}$ \\
$\text{Li}_{3}\text{BN}_{2}$ & $8.954 \times 10^{-3}$ & $8.953 \times 10^{-3}$ & $9.226 \times 10^{1}$ & $9.226 \times 10^{1}$ \\
$\text{Li}_{3}\text{PO}_{4}$ & $5.137 \times 10^{-14}$ & $-1.749 \times 10^{-10}$ & $6.544 \times 10^{1}$ & $6.544 \times 10^{1}$ \\
$\text{Li}(\text{NdTiO}_{4})$ & $4.022 \times 10^{0}$ & $4.022 \times 10^{0}$ & $5.729 \times 10^{1}$ & $5.729 \times 10^{1}$ \\
$\text{Li}_{7}\text{P}_{3}\text{S}_{11}$ & $5.212 \times 10^{1}$ & $4.892 \times 10^{1}$ & $4.159 \times 10^{0}$ & $9.593 \times 10^{-1}$ \\
$\text{Li}_{4}\text{Zn}(\text{PO}_{4})_{2}$ & $3.278 \times 10^{-1}$ & $3.278 \times 10^{-1}$ & $4.841 \times 10^{1}$ & $4.841 \times 10^{1}$ \\
$\text{Li}_{5}\text{La}_{3}\text{Ta}_{2}\text{O}_{12}$ & $5.999 \times 10^{-1}$ & $5.659 \times 10^{-1}$ & $2.788 \times 10^{1}$ & $2.785 \times 10^{1}$ \\
$\text{LiLaNb}_{2}\text{O}_{7}$ & $2.171 \times 10^{-3}$ & $2.171 \times 10^{-3}$ & $1.215 \times 10^{1}$ & $1.215 \times 10^{1}$ \\
$\text{Li}_{3}\text{BiO}_{4}$ & $8.900 \times 10^{0}$ & $8.900 \times 10^{0}$ & $3.726 \times 10^{-9}$ & $-4.063 \times 10^{-7}$ \\
$\text{LiGaGeO}_{4}$ & $1.170 \times 10^{-9}$ & $1.169 \times 10^{-9}$ & $6.209 \times 10^{0}$ & $6.209 \times 10^{0}$ \\
$\text{LiGaO}_{2}$ & $2.307 \times 10^{-13}$ & $-2.377 \times 10^{-11}$ & $2.340 \times 10^{-9}$ & $2.316 \times 10^{-9}$ \\
$\text{Li}_{2.5}\text{V}_{2}(\text{PO}_{4})_{3}$ & $4.989 \times 10^{0}$ & $4.989 \times 10^{0}$ & $1.741 \times 10^{0}$ & $1.741 \times 10^{0}$ \\
$\text{Li}_{7}\text{La}_{3}\text{Zr}_{2}\text{O}_{12}$ & $4.346 \times 10^{-1}$ & $4.330 \times 10^{-1}$ & $4.660 \times 10^{0}$ & $4.659 \times 10^{0}$ \\
$\text{Li}_{8}\text{GeP}_{4}$ & $4.248 \times 10^{0}$ & $4.230 \times 10^{0}$ & $3.930 \times 10^{-1}$ & $3.750 \times 10^{-1}$ \\
$\text{LiVO}_{3}$ & $2.464 \times 10^{0}$ & $2.464 \times 10^{0}$ & $1.621 \times 10^{0}$ & $1.621 \times 10^{0}$ \\
$\text{Li}_{7}\text{BiO}_{6}$ & $2.392 \times 10^{-1}$ & $2.376 \times 10^{-1}$ & $3.128 \times 10^{0}$ & $3.126 \times 10^{0}$ \\
$\text{Li}_{3}\text{AlN}_{2}$ & $1.518 \times 10^{0}$ & $1.518 \times 10^{0}$ & $4.023 \times 10^{-1}$ & $4.022 \times 10^{-1}$ \\
$\text{Li}_{3}\text{P}$ & $1.192 \times 10^{-4}$ & $-7.029 \times 10^{-1}$ & $1.974 \times 10^{-2}$ & $-6.833 \times 10^{-1}$ \\
$\text{Li}_{7}\text{SbO}_{6}$ & $1.604 \times 10^{-2}$ & $1.597 \times 10^{-2}$ & $9.020 \times 10^{-1}$ & $9.020 \times 10^{-1}$ \\
$\text{LiSi}_{2}\text{N}_{3}$ & $1.753 \times 10^{-9}$ & $-6.170 \times 10^{-5}$ & $7.175 \times 10^{-1}$ & $7.174 \times 10^{-1}$ \\
$\text{LiSbO}_{2}$ & $4.147 \times 10^{-6}$ & $4.147 \times 10^{-6}$ & $4.347 \times 10^{-1}$ & $4.347 \times 10^{-1}$ \\
$\text{LiZr}_{2}(\text{PO}_{4})_{3}$ & $9.658 \times 10^{-3}$ & $9.657 \times 10^{-3}$ & $3.877 \times 10^{-1}$ & $3.877 \times 10^{-1}$ \\
$\text{Li}_{4}\text{GeS}_{4}$ & $2.764 \times 10^{-3}$ & $2.564 \times 10^{-3}$ & $3.977 \times 10^{-1}$ & $3.975 \times 10^{-1}$ \\
$\text{Li}_{5}\text{BiO}_{5}$ & $3.355 \times 10^{-1}$ & $3.355 \times 10^{-1}$ & $1.162 \times 10^{-1}$ & $1.162 \times 10^{-1}$ \\
$\text{Li}(\text{LaTiO}_{4})$ & $2.111 \times 10^{-2}$ & $2.111 \times 10^{-2}$ & $2.372 \times 10^{-1}$ & $2.372 \times 10^{-1}$ \\
$\text{Li}_{6}\text{Ge}_{2}\text{O}_{7}$ & $2.384 \times 10^{-1}$ & $2.375 \times 10^{-1}$ & $6.945 \times 10^{-6}$ & $-8.431 \times 10^{-4}$ \\
$\text{Li}_{3}\text{N}$ & $1.118 \times 10^{-2}$ & $-1.973 \times 10^{-1}$ & $1.075 \times 10^{-1}$ & $-1.010 \times 10^{-1}$ \\
$\text{Li}_{1.5}\text{Al}_{0.5}\text{Ge}_{1.5}(\text{PO}_{4})_{3}$ & $2.452 \times 10^{-15}$ & $-1.900 \times 10^{-1}$ & $2.172 \times 10^{-1}$ & $2.718 \times 10^{-2}$ \\
$\text{Li}_{6}\text{CuB}_{4}\text{O}_{10}$ & $2.262 \times 10^{-6}$ & $2.262 \times 10^{-6}$ & $1.741 \times 10^{-1}$ & $1.741 \times 10^{-1}$ \\
$\text{Li}_{3}\text{VO}_{4}$ & $2.403 \times 10^{-2}$ & $2.403 \times 10^{-2}$ & $1.561 \times 10^{-1}$ & $1.561 \times 10^{-1}$ \\
$\text{Li}_{4}\text{SnS}_{4}$ & $9.709 \times 10^{-2}$ & $2.709 \times 10^{-2}$ & $1.196 \times 10^{-1}$ & $4.959 \times 10^{-2}$ \\
$\text{Li}_{2}\text{BaP}_{2}\text{O}_{7}$ & $5.041 \times 10^{-15}$ & $-1.095 \times 10^{-14}$ & $7.969 \times 10^{-2}$ & $7.969 \times 10^{-2}$ \\
$\text{LiAlH}_{4}$ & $6.092 \times 10^{-2}$ & $6.091 \times 10^{-2}$ & $3.167 \times 10^{-6}$ & $-5.533 \times 10^{-6}$ \\
$\text{Li}_{6}\text{Zr}_{2}\text{O}_{7}$ & $5.996 \times 10^{-2}$ & $5.996 \times 10^{-2}$ & $1.522 \times 10^{-6}$ & $1.002 \times 10^{-6}$ \\
$\text{LiGd}(\text{PO}_{3})_{4}$ & $4.814 \times 10^{-2}$ & $4.814 \times 10^{-2}$ & $6.627 \times 10^{-13}$ & $-3.373 \times 10^{-13}$ \\
$\text{Li}_{3}\text{Na}_{5}(\text{TiS}_{4})_{2}$ & $8.293 \times 10^{-12}$ & $-8.800 \times 10^{-3}$ & $8.001 \times 10^{-10}$ & $-8.800 \times 10^{-3}$ \\
$\text{LiGa}_{0.5}\text{Al}_{0.5}\text{GeO}_{4}$ & $1.054 \times 10^{-2}$ & $1.054 \times 10^{-2}$ & $4.284 \times 10^{-15}$ & $-9.957 \times 10^{-13}$ \\
$\text{Li}_{4}\text{SnSe}_{4}$ & $7.891 \times 10^{-17}$ & $-4.500 \times 10^{-5}$ & $1.169 \times 10^{-2}$ & $1.164 \times 10^{-2}$ \\
$\text{Li}_{2}\text{P}_{2}\text{S}_{6}$ & $3.604 \times 10^{-3}$ & $3.604 \times 10^{-3}$ & $2.564 \times 10^{-6}$ & $2.486 \times 10^{-6}$ \\
$\text{Li}_{2}\text{NaBP}_{2}\text{O}_{8}$ & $1.073 \times 10^{-3}$ & $1.073 \times 10^{-3}$ & $3.262 \times 10^{-18}$ & $-4.397 \times 10^{-15}$ \\
$\text{Li}_{7}\text{La}_{3}\text{Hf}_{2}\text{O}_{12}$ & $5.444 \times 10^{-16}$ & $-9.850 \times 10^{-4}$ & $7.581 \times 10^{-10}$ & $-9.850 \times 10^{-4}$ \\
$\text{LiGe}_{2}(\text{PO}_{4})_{3}$ & $2.154 \times 10^{-6}$ & $-3.308 \times 10^{-4}$ & $1.386 \times 10^{-18}$ & $-3.330 \times 10^{-4}$ \\
$\text{LiBiO}_{2}$ & $3.518 \times 10^{-4}$ & $3.138 \times 10^{-4}$ & $1.500 \times 10^{-5}$ & $-2.300 \times 10^{-5}$ \\
$\text{LiZnPS}_{4}$ & $1.700 \times 10^{-17}$ & $-5.400 \times 10^{-5}$ & $1.420 \times 10^{-18}$ & $-5.400 \times 10^{-5}$ \\
$(\text{Li}_{0.5}\text{Ce}_{0.5})(\text{MoO}_{4})$ & $6.428 \times 10^{-10}$ & $-1.300 \times 10^{-5}$ & $7.372 \times 10^{-7}$ & $-1.226 \times 10^{-5}$ \\
$\text{LiBO}_{2}$ & $2.796 \times 10^{-9}$ & $-9.997 \times 10^{-6}$ & $2.746 \times 10^{-9}$ & $-9.997 \times 10^{-6}$ \\
$\text{Li}_{3}\text{BiO}_{3}$ & $6.956 \times 10^{-6}$ & $6.716 \times 10^{-6}$ & $4.819 \times 10^{-12}$ & $-2.400 \times 10^{-7}$ \\
$\text{Li}_{4}\text{GeO}_{4}$ & $6.370 \times 10^{-6}$ & $6.368 \times 10^{-6}$ & $1.884 \times 10^{-3}$ & $1.884 \times 10^{-3}$ \\
$\text{Li}_{3}\text{SbS}_{3}$ & $6.720 \times 10^{-6}$ & $5.720 \times 10^{-6}$ & $1.610 \times 10^{-20}$ & $-1.000 \times 10^{-6}$ \\
$\text{Li}_{5}\text{GaO}_{4}$ & $5.362 \times 10^{-9}$ & $-4.995 \times 10^{-6}$ & $5.339 \times 10^{-9}$ & $-4.995 \times 10^{-6}$ \\
$\text{Li}_{3}\text{SbS}_{4}$ & $1.743 \times 10^{-9}$ & $-4.798 \times 10^{-6}$ & $1.787 \times 10^{-9}$ & $-4.798 \times 10^{-6}$ \\
$\text{Li}_{5}\text{AlS}_{4}$ & $2.266 \times 10^{-16}$ & $-1.400 \times 10^{-6}$ & $3.418 \times 10^{-15}$ & $-1.400 \times 10^{-6}$ \\
$\text{Li}_{3.5}\text{Zn}_{0.5}\text{Ga}_{0.5}(\text{PO}_{4})_{2}$ & $8.996 \times 10^{-16}$ & $-9.991 \times 10^{-13}$ & $9.380 \times 10^{-8}$ & $9.380 \times 10^{-8}$ \\
$\text{LaLiO}_{2}$ & $1.969 \times 10^{-9}$ & $1.968 \times 10^{-9}$ & $1.975 \times 10^{-9}$ & $1.974 \times 10^{-9}$ \\
$\text{Li}_{2}\text{Sr}_{2}\text{Al}(\text{PO}_{4})_{3}$ & $2.456 \times 10^{-15}$ & $-9.975 \times 10^{-13}$ & $3.088 \times 10^{-15}$ & $-9.969 \times 10^{-13}$ \\
$\text{Fe}_{2}\text{Na}_{2}\text{K}(\text{Li}_{3}\text{Si}_{12}\text{O}_{30})$ & $1.101 \times 10^{-15}$ & $-9.989 \times 10^{-13}$ & $1.429 \times 10^{-14}$ & $-9.857 \times 10^{-13}$ \\
\hline
\end{tabular}
\end{table}

\clearpage

\end{document}